\documentclass[journal]{IEEEtran}
\usepackage{amsmath,amsfonts}
\usepackage{algorithmic}
\usepackage{algorithm}
\usepackage{array}
\usepackage{textcomp}
\usepackage{stfloats}
\usepackage{url}
\usepackage{verbatim}
\usepackage{graphicx,subfigure}
\usepackage{amssymb}
\usepackage{cite}
\usepackage{amsthm} 
\usepackage{color}
\usepackage{diagbox,multirow,makecell}

\usepackage{epstopdf}
\newtheorem{lemma}{Lemma}

\newtheorem{theorem}{Theorem}

\definecolor{mygreen}{rgb}{0, 0.63, 0.32}
\definecolor{mybrown}{rgb}{0.7, 0.3, 0.0}

\begin{document}

\title{Optimal Waveform Design for Continuous Aperture Array (CAPA)-aided ISAC Systems}

\author{Junjie~Ye, \IEEEmembership{Graduate Student Member,~IEEE,} Zhaolin~Wang, \IEEEmembership{Member,~IEEE,} Yuanwei~Liu, \IEEEmembership{Fellow,~IEEE,} 
Peichang~Zhang, \IEEEmembership{Member,~IEEE,} Lei~Huang, \IEEEmembership{Senior Member,~IEEE,}
Arumugam~Nallanathan, \IEEEmembership{Fellow,~IEEE} 
\thanks{This work has been submitted to the IEEE for possible publication.  Copyright may be transferred without notice, after which this version may no longer be accessible.}
\thanks{J. Ye,  P. Zhang and L. Huang are with State Key Laboratory of Radio Frequency Heterogeneous Integration, Shenzhen University, Shenzhen, China. (e-mail: 2152432003@email.szu.edu.cn; $\lbrace$pzhang, lhuang$\rbrace$@szu.edu.cn.) \\ 
\indent Z. Wang and Y. Liu are with the Department of Electrical and Electronic Engineering, the University of Hong Kong, Hong Kong, China. (e-mail: $\{$zhaolin.wang,yuanwei$\}$@hku.hk).\\
\indent A. Nallanathan is with the School of Electronic Engineering and Computer Science, Queen Mary University of London, London, UK. (e-mail: a.nallanathan@qmul.ac.uk).}}


\maketitle

\begin{abstract}
A novel continuous-aperture-array (CAPA)-aided integrated sensing and communication (ISAC) framework is proposed. Specifically, an optimal continuous ISAC waveform is designed to form a directive beampattern for multi-target sensing while suppressing the multi-user interference (MUI). To achieve the goal of optimal waveform design, the directional beampattern of CAPA is first derived based on Green’s function, whereafter a reference sensing waveform is obtained through wavenumber-domain optimization. Based on the reference sensing waveform, a weighted functional programming on the tradeoff between sensing beampattern mismatch and MUI is formulated. To solve the resulting problem, an optimal CAPA-ISAC waveform structure is analytically derived using a Lagrangian-transformation and calculus-of-variations method, where the Lagrangian multiplier associated with the optimal waveform structure is determined via Bisection search. The obtained optimal waveform reveals that it is concurrently affected by the reference sensing waveform, the channel correlations and the channel-symbol correlations.  Finally, numerical results validate the effectiveness of the proposed system and waveform design, demonstrating that CAPA can achieve significant performance gains against the  ISAC designs based on conventional spatially discrete array in both sensing accuracy and communication reliability.
\end{abstract}


\section{Introduction}
\IEEEPARstart{I}{ntegrated} sensing and communication (ISAC) has emerged as one of the most promising techniques for next-generation wireless networks, attracting considerable attention from both academia and industry. Specifically, ISAC intends to unify communication and sensing functionalities by sharing spectrum resources, signal processing techniques, and hardware infrastructures \cite{Liufan_ISAC_survey}. In the early stages of ISAC development, the primary focus was spectrum sharing, enabling communication and sensing systems to coexist within the same frequency band with mitigated mutual interference. For example, \cite{Liufan_Coexistence} investigated the coexistence of downlink multiple-input multiple-output (MIMO) communication and MIMO radar, and proposed a robust beamforming approach for interference suppression. Similarly, \cite{Rihan_spectrum_codesign} developed a two-tier alternating-optimization framework for spectrum sharing, achieving interference alignment between the two systems and significantly reducing the impact of mutual interference.

Considering the architectural similarity between sensing and communication transceivers, the research has gradually shifted to deeper-level integrations, where both functionalities share not only spectrum but also waveform and signal processing designs. For instance, \cite{LiuFan_OptimalWaveform} proposed an integrated platform that transmitted a dedicated ISAC waveform to simultaneously mitigate multi-user interference (MUI) and minimize sensing beampattern mismatch. In \cite{LiuXiang_ISAC}, a unified beamforming framework was developed to jointly optimize transmit beamformers for radar waveforms and communication symbols. Beyond waveform design, the joint signal processing of communication and sensing has also been extensively investigated \cite{Zhangjian_ISAC_SP}. For instance, \cite{Zhangpin_MUSIC_ISAC} applied multiple signal classification for high-accuracy target parameter estimation, while \cite{ISAC_Estimation_Tensor} introduced a tensor-based approach to simultaneously address channel estimation and target sensing. Moreover, ISAC has been applied in diverse scenarios, including energy-efficient designs for power-constrained systems \cite{Tony_EE,JJ_EE,Derrick_EE}, robust designs under dynamic environments \cite{Chang_robust,LiYong_Robust,LiYe_robust}, and data-driven approaches for computation-intensive tasks \cite{LiuRang_SLP_learning,JJ_Learning,Chunguo_DL}.

Despite these advances, most of the aforementioned studies relied on conventional MIMO array architectures, whose physical limitations inherently restricted communication rate and sensing accuracy. To overcome this issue, research has evolved towards large-scale MIMO systems, including massive MIMO (mMIMO) and extremely large-scale MIMO (XL-MIMO) arrays \cite{Jiayi_XLMIMO}. Specifically, these systems typically deploy hundreds or even thousands of antennas at the base station (BS), leading to channel hardening phenomena \cite{XL_MIMO_Zhilong}. Consequently, more favorable propagation conditions can be established, which contributes to reducing inter-cell interference and enhancing communication reliability \cite{Qingchao_XLMIMO}. Meanwhile, large-scale MIMO arrays can also improve sensing performance, as the increased number of antennas enables finer angular resolution and narrower spatial beam \cite{Zhaolin_Large_Aperture}.

Motivated by these advantages, numerous studies have explored the potential of large-scale MIMO for ISAC. For example, \cite{Zengyong_mMIMO_ISAC} implemented a large-scale antenna array in an ISAC node and proposed a codebook-based beamforming strategy to balance beamforming gain and time overhead. In \cite{ISAC_mMIMO_Qihao}, computation latency was considered in a mMIMO-enabled joint sensing, communication, and computation system, where beamforming and task allocation were optimized. To reduce the burden of processing high-dimensional data in mMIMO systems, \cite{Gaozhen_mMIMO_ISAC} introduced a compressed-sampling-based signal processing framework. Furthermore, some research has extended ISAC to the XL-MIMO regime. In \cite{Zhangrui_XLMIMO_ISAC} and \cite{Zhaolin_XLMIMO_ISAC}, the authors exploited the low-rank structure of channel matrices to efficiently solve rank-constrained joint beamforming problems. Nevertheless, these MIMO systems still relied on independent RF chains and were inherently constrained by half-wavelength element spacing. To overcome such limitations, the concept of reconfigurable holographic surfaces (RHSs) has been proposed \cite{Boya_HMIMO,Boya_HMIMO_Mutual,NiWei_ISAC_RHS}. By employing parallel-plate waveguides, RHSs generate directional beams by converting guided waves into leaky waves. However, the current implementations of RHS remain limited to a finite number of radiating elements.

In these developments, the core idea was to integrate an increasing number of antennas within a limited surface area to approach the ultimate spatial degrees of freedom (DoFs) and channel capacity limits. This trend naturally leads to the concept of a continuous electromagnetic (EM) aperture, known as the continuous aperture array (CAPA)  \cite{Yuanwei_CAPA}. Recent studies have conducted fundamental investigations of CAPA, aiming to characterize its achievable DoFs \cite{Nallan_CAP_Performance} and capacity bounds \cite{Jensen_CAP_EM_Channel}. Building on these theoretical foundations, several practical schemes have been explored for CAPA-aided communication systems. For instance, \cite{Luca_Wavenumber} proposed a wavenumber-division multiplexing scheme to mitigate MUI in communication systems, while \cite{Zeyu_CAHMIMO} analyzed the signal and interference channels to identify dominant interference sources. Beyond interference mitigation, some works focused on waveform and resource optimization. Notably, Fourier-based methods \cite{linglong_CAPA,Linglong_CAPA_VS_SPDA} and functional programming approaches \cite{Zhaolin_CAPA_BF,Zhaolin_CAPA_Optimal_BF} have been developed for transmit current pattern design. In addition to communication, CAPA has demonstrated significant potential for sensing. In \cite{jianghao_CAPA_Sensing},  a maximum-likelihood estimator for CAPA-based target detection was proposed, and the sensing Cramér–Rao bound (CRB) was optimized via a subspace-based approach. More recently, a few studies have incorporated CAPA into ISAC systems. In \cite{Ouyang_CAPA_ISAC2}, CAPA-aided ISAC was analyzed for both single-user single-target and multi-user multi-target scenarios, where a Fourier-based beamforming method was developed. Subsequently, \cite{Ouyang_CAPA_ISAC} investigated the rate–CRB tradeoff, while \cite{Boqun_DLUL_CAPA_ISAC} studied CAPA-aided ISAC systems from an information-theoretic perspective, revealing achievable rate gains in uplink and downlink scenarios. 

However, the above studies on CAPA-aided ISAC systems mainly focused on task-dependent metrics under specific models, which motivated us to develop a intuitive and task-independent framework over the spatial energy distribution control. Specifically, we aim to design an optimal transmit  waveform for simultaneous beampattern shaping and MUI suppression. In sensing, the transmit beampattern plays a critical role in multi-target sensing scenarios, as it determines the spatial energy distribution and directly affects targets' illuminations. In communication, effective MUI mitigation are essential tasks for accurate symbol recovery.  The main contributions are summarized as follows:

\begin{itemize}
    \item  A novel CAPA-aided ISAC framework is proposed and an optimal transmit waveform design  for multi-target sensing
 and multi-user communication. Specifically, the transmit source current patterns, the communication MUI energy, and the sensing beampattern mismatch are first characterized. Based on these models, a functional optimization problem is formulated to jointly minimize sensing beampattern mismatch and MUI energy.   
    
    \item To obtain the ideal current source pattern for sensing reference, we derive  EM power and directional beampattern based on the Green's function. Subsequently, a minimal beam gain maximization problem is formulated for multi-target illumination, which is solved via a wavenumber-domain optimization.
    
    \item  To tackle the functional problem for the ISAC waveform design, we propose a calculus of variations (CoV)-based algorithm. First, the original problem is converted into an unconstrained optimization via Lagrangian transformation, whereafter the optimal waveform structure for ISAC is derived by employing CoV. Then, the Lagrangian multiplier associated with the optimal waveform structure is obtained via the Bisection search. The obtained solution reveals its relationship with the reference sensing waveform, the channels and the communication symbols.

    \item Comprehensive numerical results are presented to demonstrate the effectiveness of the proposed framework.  The results reveal that CAPA can achieve significant gains over the spatially discrete array (SPDA) in terms of both sensing beampattern and communication bit error rate (BER).  Moreover, the tradeoff between sensing and communication can be  improved by enlarging the aperture sizes as well as increasing the carrier frequencies.

\end{itemize}

The remainder of this paper is organized as follows. Section \ref{sec:model} presents the architecture of the CAPA-aided ISAC system and formulates the joint sensing–communication optimization problem. Section \ref{sec:EM_ref} derives the directional beampattern for CAPA and obtains the reference current source pattern for sensing. Section \ref{algorithm} details the proposed CoV-based algorithm and analyzes its computational complexity. Section \ref{simulation} provides numerical results, and Section \ref{conclusions} concludes the paper.

\textit{Notation}: Regular, bold lowercase, and uppercase letters denote scalars, vectors, and matrices, respectively. The symbols $\mathbb{C}^{M\times N}$ and $\mathbb{R}^{M\times N}$ represent complex and real space with a dimension of $M\times N$. Moreover, $(\cdot)^{-1}$, $(\cdot)^\mathrm{H}$, $(\cdot)^\mathrm{T}$, and $(\cdot)^*$ denote the inverse, conjugate transpose, transpose, and conjugate operations, respectively. $\int_{\mathcal{S}} f(\mathbf{s}) d\mathbf{s}$ represents the integral of function $f(\mathbf{s})$ over the field $\mathcal{S}$. The ceiling operator is denoted by $\lceil \cdot \rceil$, while $|\cdot|$ and $|\cdot|$ represent the absolute value and norm, respectively. Finally, $\mathfrak{R}\{x\}$ denotes the real part of $x$, and $\jmath$ is the imaginary unit.

\section{System Model} \label{sec:model}
\begin{figure}
    \centering
    \includegraphics[width=0.85\linewidth]{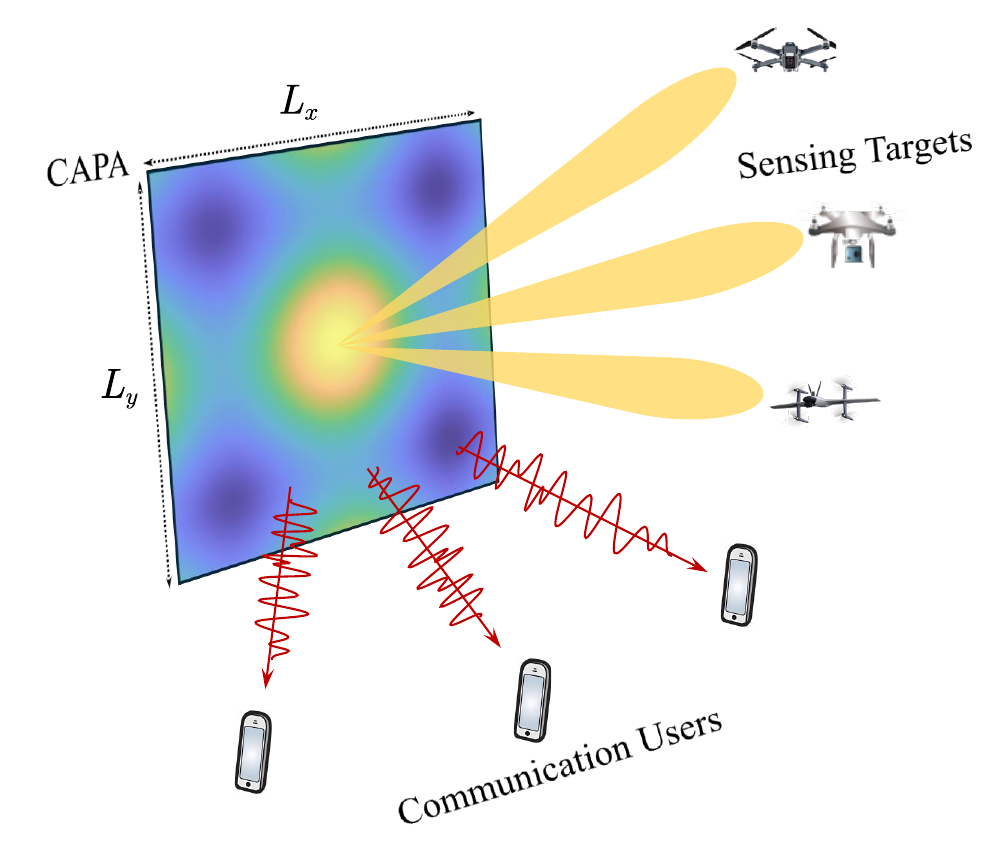}
    \caption{A scenario of a CAPA-aided ISAC system that simultaneously senses multiple targets and serves multiple users.}
    \label{fig:system}
\end{figure}

As illustrated in Fig. \ref{fig:system}, we consider a CAPA-aided ISAC system, where the CAPA transmitter is deployed at the BS to simultaneously support multi-user communication and multi-target sensing. In this system, $K$ communication users are served, each equipped with a single uni-polarized receive antenna, while $T$ point targets located in distinct directions are sensed concurrently. 

\subsection{Transmit Model}
The CAPA transmitter is a planar continuous radiation surface $\mathcal{S}_\mathrm{T}$, where the lengths along the $x$- and $y$-axes are denoted as $L_x$ and $L_y$, respectively. The total area of the CAPA is given by $|\mathcal{S}_\mathrm{T}| = L_x \cdot L_y = A_\mathrm{T}$. The CAPA is capable of generating controllable source currents to radiate EM waves for both communication and sensing purposes. Without loss of generality, the center of the CAPA is placed at the origin, and the coordinate of an arbitrary point on the surface is denoted by $\mathbf{s} = \left[s_x, s_y, 0\right]^\mathrm{T} \in \mathcal{S}_\mathrm{T}$. In the considered system, narrowband transmission is assumed, and the source current density at point $\mathbf{s}$ is denoted by $\mathbf{j}(\mathbf{s})$. In general, the source current density is tri-polarized along three orthogonal directions, i.e.,
\begin{align}
     \mathbf{j}(\mathbf{s})=j_x(\mathbf{s})\mathbf{u}_x + j_y(\mathbf{s})\mathbf{u}_y + j_z(\mathbf{s})\mathbf{u}_z \in \mathbb{C}^{3\times 1},
\end{align}
where $\mathbf{u}_x, \mathbf{u}_y, \mathbf{u}_z \in \mathbb{R}^{3 \times 1}$ are unit orthogonal vectors, and $j_x(\mathbf{s})$, $j_y(\mathbf{s})$, and $j_z(\mathbf{s})$ denote the current components along the orthogonal directions, respectively. In this study, we assume that the source current is uni-polarized along the $y$-axis. Accordingly, the source current density can be expressed as
\begin{align}
     \mathbf{j}(\mathbf{s}) \triangleq j(\mathbf{s})\mathbf{u}_y,
\end{align}
where $\mathbf{u}_y=[0,1,0]^\mathrm{T}$ and $j(\mathbf{s}) \triangleq j_y(\mathbf{s})$ is introduced for notational simplicity. According to \cite{Luca_Wavenumber}, the energy of $\mathbf{j}(\mathbf{s})$ is given by
\begin{align} 
    P_0=\int_{\mathcal{S}_\mathrm{T}} \left|j(\mathbf{s})\right|^2 d\mathbf{s},
\end{align}
which serves as the upper bound of the physical EM radiation power. Accordingly, imposing an constraint on $P_0$ can effectively limit the radiated power, i.e, $P_0\leq P_t$ with $P_t$ representing the power budget.

\subsection{Communication Model}
Let $\mathbf{r}_k$ denote the location of the communication user $k$. According to Maxwell's equations, the electric field at user $k$ induced by $\mathbf{j}(\mathbf{s})$ can be expressed as \cite{Zhaolin_CAPA_BF}
\begin{align}    \label{electric_field}
\mathbf{e}_k=\int_{\mathcal{S}_\mathrm{T}} \mathbf{G}(\mathbf{r}_k,\mathbf{s})\mathbf{j}(\mathbf{s}) d\mathbf{s} \in \mathbb{C}^{3\times 1}, ~\forall k,
\end{align}
where $\mathbf{G}(\mathbf{r}_k, \mathbf{s})$ denotes an integral kernel function. In line-of-sight (LoS) scenarios, $\mathbf{G}(\mathbf{r}_k, \mathbf{s})$ corresponds to the Green’s function. For conciseness, a detailed expression of $\mathbf{G}(\mathbf{r}_k, \mathbf{s})$ is provided in Section \ref{subsec:Green}.

For user $k$, a uni-polarized receive antenna is equipped, with polarization direction denoted by $\mathbf{u}_k \in \mathbb{R}^{3\times 1}$. Accordingly, the received signal at  user $k$ can be expressed as
\begin{align}
    y_k &= \mathbf{u}_k^\mathrm{T}\mathbf{e}_k + n_k \notag\\
    &= \int_{\mathcal{S}_\mathrm{T}} \mathbf{u}_k^\mathrm{T}\mathbf{G}(\mathbf{r}_k,\mathbf{s})\mathbf{u}_y  j(\mathbf{s}) d\mathbf{s} + n_k \notag \\
    &= \int_{\mathcal{S}_\mathrm{T}} H_k(\mathbf{s})  j(\mathbf{s}) d\mathbf{s} + n_k,~\forall k,
\end{align}
where $H_k(\mathbf{s})=\mathbf{u}_k^\mathrm{T}\mathbf{G}(\mathbf{r}_k,\mathbf{s})\mathbf{u}_y$ is the equivalent communication channel, and $n_k \in \mathbb{C}$ is the additive white Gaussian noise with zero mean and $\sigma_k^2$ variance.

Let $c_k$ be the desired constellation symbol for user $k$. Accordingly, the received signals can be rewritten as
\begin{align} \label{decode_receive_model}
    y_k &= c_k + \underbrace{\left( \int_{\mathcal{S}_\mathrm{T}} H_k(\mathbf{s})  j(\mathbf{s}) d\mathbf{s} - c_k \right)}_{\text{MUI}}+ n_k,~\forall k,
\end{align}
where the second term can be recognized as the MUI for  user $k$. The  energy associated with MUI is given by
\begin{align}
    \overline{P}_{c,k}= \left|\int_{\mathcal{S}_\mathrm{T}} H_k(\mathbf{s})  j(\mathbf{s}) d\mathbf{s} - c_k \right|^2,~\forall k.
\end{align}
It has been shown in \cite{LiuFan_OptimalWaveform} that the energy of the MUI signal directly affects the SINR of the downlink users. Given a constellation symbol with fixed energy, the SINR increases as the MUI energy reduces. Therefore, we adopt the sum of the MUI energy of all the users as the communication metric, which is given by 
\begin{align}\label{MUI}
    f_c =\sum_{k=1}^K \overline{P}_{c,k}.
\end{align}

\subsection{Sensing Model}
For sensing purposes, an effective way is to design the transmit beampattern where the beams are pointed towards the target directions.  Given an ideal waveform $j_d(\mathbf{s})$ that corresponds to a well-designed sensing beampattern,  $j(\mathbf{s})$ can be optimized to approach $j_d(\mathbf{s})$. As a result, we adopt the difference between  $j(\mathbf{s})$ and  $j_{d}(\mathbf{s})$ as the sensing metric, which can be characterized by
\begin{align} \label{Mismatch}
    f_s = \int_{\mathcal{S}_\mathrm{T}}   \left|  j(\mathbf{s}) - j_{d}(\mathbf{s}) \right|^2 d\mathbf{s} .
\end{align}
Here, the difference is measured using a functional norm defined in the Banach space, since $j(\mathbf{s})$ and $j_d(\mathbf{s})$ are functions with respect to (w.r.t.) $\mathbf{s}$.  Furthermore, the procedure for obtaining 
 $j_d(\mathbf{s})$ is elaborated in Section \ref{sec:EM_ref}.

\subsection{Problem Formulation}
To concurrently guarantee the sensing and communication performance,  a joint sensing beampattern mismatch and the multi-user interference minimization problem is formulated,  where an optimal current waveform $j(\mathbf{s})$ is  designed. Accordingly, the optimization problem can be given by
\begin{subequations} \label{ISAC_formulation}
\begin{align}
    \min_{j(\mathbf{s})} &~~ f_{s,c}(j(\mathbf{s}))=\rho \cdot f_c (j(\mathbf{s})) + (1-\rho) \cdot f_r(j(\mathbf{s}))  \\
    \mathrm{s.t.} &~~  \int_{\mathcal{S}_\mathrm{T}} \left|j(\mathbf{s})\right|^2 d\mathbf{s}= P_t,
\end{align} 
\end{subequations}
where $0\leq \rho \leq 1$ is a weighting coefficient that controls the trade-off between sensing and communication performance. Besides, the constraint enforces a finite transmit power $P_t$ for the CAPA transmitter.  Considering that the sensing system is generally required to transmit at its maximum power, an equality constraint is enforced on the total power budget \cite{LiuFan_OptimalWaveform}. The problem (\ref{ISAC_formulation}) is a functional programming problem, where the optimization variable is a continuous function. Consequently, conventional waveform optimization methods developed for discrete-array systems are not directly applicable. It is worth noting that this formulation represents a trade-off design between sensing and communication performance, and the obtained solution can reach a Pareto-optimal point \cite{LiuFan_OptimalWaveform}.

\section{Optimal EM Waveform Design for Sensing} \label{sec:EM_ref}
In this section, an optimal EM waveform for sensing in the CAPA-aided system is designed as the reference sensing waveform $j_d(\mathbf{s})$. For analytical clarity, we focus on the target illumination in angular domain, where the Green’s function is simplified to a far-field case. Subsequently, the directional beampattern expression is derived, based on which the reference sensing waveform is designed via Fourier-based approach.

\subsection{Far-Field Approximation of Green's Function} \label{subsec:Green}
  We denote $\mathbf{r}$ as an arbitrary location in the radiation space. Based on Maxwell's equations, the electric field response at  $\mathbf{r}$ is given in (\ref{electric_field}). Considering LoS propagation in unbounded and homogeneous mediums, the channel function $\mathbf{G}(\mathbf{r},\mathbf{s})$ can be expressed as \cite{linglong_CAPA}
\begin{align} \label{green_func}
    \mathbf{G}(\mathbf{r},\mathbf{s})=-\frac{\jmath\eta e^{-\jmath\frac{2\pi}{\lambda}\|\mathbf{r}-\mathbf{s}\|}}{2\lambda\|\mathbf{r}-\mathbf{s}\|}\left(\mathbf{I}_3-\frac{(\mathbf{r}-\mathbf{s})(\mathbf{r}-\mathbf{s})^\mathrm{T}}{\|\mathbf{r}-\mathbf{s}\|^2}\right),
\end{align}
where $\eta$ represents the intrinsic impedance of EM wave propagation in free space, and $\lambda$ denotes the wavelength of the EM wave. While $\mathbf{G}(\mathbf{r},\mathbf{s})$ capture the full near-field behaviour, we primarily focus on far-field approximations and the radiation in angular domain. First, $\mathbf{r}$ can be parameterized as
\begin{align}
    \mathbf{r}=r\mathbf{k}(\theta,\phi)=r\left[ \cos \theta \sin\phi, \sin \theta\sin\phi, \cos\phi\right],
\end{align}
where $r$ denotes the distance from the CAPA center to $\mathbf{r}$, and $\mathbf{k}(\theta,\phi)$ specifies the  propagation direction vector with azimuth angle $\theta$ and elevation angle $\phi$. In the far-field regime, the distance term $\|\mathbf{r}-\mathbf{s}\|$ can be approximated as \cite{Yuanwei_Nearfield}
\begin{align} \label{near_field_approx}
    \|\mathbf{r}-\mathbf{s}\| &= \|r\mathbf{k}(\theta,\phi)-\mathbf{s}\|\notag\\
    &=\sqrt{r^2-2r{\mathbf{k}^\mathrm{T}(\theta,\phi)\mathbf{s}} + {\|\mathbf{s}\|^2}} \notag \\
    & \approx r - \mathbf{k}^\mathrm{T}(\theta,\phi)\mathbf{s},
\end{align} 
where this approximation neglects the higher-order terms w.r.t $r$ when the  distance $r$ is sufficiently large compared to the aperture size.
By substituting the approximation in (\ref{near_field_approx}) into (\ref{green_func}), the far-field Green’s function can be rewritten as
\begin{align}\label{green_approx}
    \mathbf{G}(\mathbf{r},\mathbf{s})&\approx -\frac{\jmath\eta e^{-\jmath\frac{2\pi}{\lambda}\left(r - \mathbf{k}^\mathrm{T}(\theta,\phi)\mathbf{s}\right)}}{2\lambda\left(r - \mathbf{k}^\mathrm{T}(\theta,\phi)\mathbf{s}\right)}\left(\mathbf{I}_3-\hat{\mathbf{p}}\hat{\mathbf{p}}^\mathrm{T}\right) \notag\\
    &= \frac{-\jmath\eta e^{-\jmath\frac{2\pi}{\lambda}r } \cdot e^{\jmath\frac{2\pi}{\lambda}{\mathbf{k}^\mathrm{T}(\theta,\phi)\mathbf{s}}}}{2\lambda\left(r - \mathbf{k}^\mathrm{T}(\theta,\phi)\mathbf{s}\right)} \left(\mathbf{I}_3-\hat{\mathbf{p}}\hat{\mathbf{p}}^\mathrm{T}\right)\notag\\
    &\approx  \frac{-\jmath\eta e^{-\jmath\frac{2\pi}{\lambda}r } \cdot e^{\jmath\frac{2\pi}{\lambda}{\mathbf{k}^\mathrm{T}(\theta,\phi)\mathbf{s}}}}{2\lambda r } \left(\mathbf{I}_3-\hat{\mathbf{r}}\hat{\mathbf{r}}^\mathrm{T}\right),
\end{align}    
where $\hat{\mathbf{p}}=\frac{\mathbf{r}-\mathbf{s}}{\|\mathbf{r}-\mathbf{s}\|}$ and  $\hat{\mathbf{r}}=\frac{\mathbf{r}}{\|\mathbf{r}\|}$.

\subsection{EM Power and Directional Beampattern}
 Given  the electric field response at $\mathbf{r}$ in (\ref{electric_field}),  the power density at this location is proportional to \cite{Luca_Wavenumber}
\begin{align} \label{bp_power}
    P(\mathbf{r})& \propto  \mathbf{e}^\mathrm{H}(\mathbf{r})\mathbf{e}(\mathbf{r})=\left\|\int_{\mathcal{S}_\mathrm{T}}  \mathbf{G}(\mathbf{r},\mathbf{s}) \mathbf{j}(\mathbf{s}) d\mathbf{s}\right\|^2.
\end{align}   
 By substituting (\ref{green_approx}) into (\ref{bp_power}), the far-field power density can be obtained as
 \begin{align}     \label{bp_power_approx}
 &P(r,\theta,\phi)  \notag \\
    \approx & \left\|\int_{\mathcal{S}_\mathrm{T}} \! \frac{-\jmath\eta  \!\cdot\! e^{-\jmath\frac{2\pi}{\lambda}r }\!\cdot\! e^{\jmath\frac{2\pi}{\lambda}{\mathbf{k}^\mathrm{T}(\theta,\phi)\mathbf{s}}}}{2\lambda r} \left(\mathbf{I}_3\!-\!\hat{\mathbf{r}}\hat{\mathbf{r}}^\mathrm{T}\right) \mathbf{j}(\mathbf{s}) d\mathbf{s}\right\|^2 \notag\\
    =& \frac{-\eta^2 }{4\lambda^2 r^2}\left\|\int_{\mathcal{S}_\mathrm{T}}  { e^{\jmath\frac{2\pi}{\lambda}{\mathbf{k}^\mathrm{T}(\theta,\phi)\mathbf{s}}}}\left(\mathbf{I}_3-\hat{\mathbf{r}}\hat{\mathbf{r}}^\mathrm{T}\right)\mathbf{u}_y j(\mathbf{s}) d\mathbf{s}\right\|^2 \notag\\
    =& ~\gamma(r) \cdot A(\theta,\phi),
\end{align}    
where $e^{-\jmath\frac{2\pi}{\lambda}r }$ is dropped since $|e^{-\jmath\frac{2\pi}{\lambda}r }|^2=1$. In (\ref{bp_power_approx}), $\gamma(r)=\frac{-\eta^2 }{4\lambda^2 r^2}$ is  a distant-related path-loss coefficient, while $A(\theta,\phi)=\left\|\int_{\mathcal{S}_\mathrm{T}}  { e^{\jmath\frac{2\pi}{\lambda}{\mathbf{k}^\mathrm{T}(\theta,\phi)\mathbf{s}}}}\left(\mathbf{I}_3-\hat{\mathbf{r}}\hat{\mathbf{r}}^\mathrm{T}\right)\mathbf{u}_y j(\mathbf{s}) d\mathbf{s}\right\|^2$ denotes the beam gain in the direction of ($\theta$, $\phi$), which can be further simplified as 
  \begin{align}\label{bp_farfield}
    A(\theta,\phi)=&\left\|\int_{\mathcal{S}_\mathrm{T}}  e^{\jmath\frac{2\pi}{\lambda}{\mathbf{k}^\mathrm{T}(\theta,\phi)\mathbf{s}}}\left(\mathbf{I}_3-\hat{\mathbf{r}}\hat{\mathbf{r}}^\mathrm{T}\right)\mathbf{u}_y j(\mathbf{s}) d\mathbf{s}\right\|^2 \notag\\
    =&\left|\int_{\mathcal{S}_\mathrm{T}}  e^{\jmath\frac{2\pi}{\lambda}{\mathbf{k}^\mathrm{T}(\theta,\phi)\mathbf{s}}} j(\mathbf{s}) d\mathbf{s}\right|^2 \cdot \left\|\left(\mathbf{I}_3-\hat{\mathbf{r}}\hat{\mathbf{r}}^\mathrm{T}\right)\mathbf{u}_y\right\|^2 \notag \\
    =& \xi^2(\theta,\phi) \left|\int_{\mathcal{S}_\mathrm{T}}  a(\theta,\phi,\mathbf{s}) j(\mathbf{s}) d\mathbf{s}\right|^2. 
\end{align}  
In (\ref{bp_farfield}), $\xi(\theta,\phi)=\left\|\left(\mathbf{I}_3-\hat{\mathbf{r}}\hat{\mathbf{r}}^\mathrm{T}\right)\mathbf{u}_y\right\|$ is the directional beam gain coefficient, while $a(\theta,\phi,\mathbf{s})=e^{\jmath\frac{2\pi}{\lambda}{\mathbf{k}^\mathrm{T}(\theta,\phi)\mathbf{s}}}$ is the steering function w.r.t the point $\mathbf{s}$  in the CAPA surface.

\subsection{Reference Sensing Waveform Design} \label{subsec:reference_signal}
We denote the target directions as $\mathrm{\Theta}=\{(\theta_1,\phi_1),(\theta_2,\phi_2),\cdots,(\theta_T,\phi_T)\}$. 
To ensure reliable sensing performance, the reference sensing waveform $j_d(\mathbf{s})$ should be designed to concentrate the radiated energy towards all the target directions, thereby adequately illuminating the targets for effective detection and accurate estimation. Specifically, we formulate the design of $j_d(\mathbf{s})$ as a max–min beam gain optimization problem, which maximizes the minimal beam gain across all target directions:
\begin{subequations} \label{ref_design_origin}
\begin{align}
    \max_{j_d(\mathbf{s})} &~~ \min_{\{\theta_l,\phi_l\} \in \mathrm{\Theta}} ~~\xi^2(\theta_l,\phi_l) \left|\int_{\mathcal{S}_\mathrm{T}}  a(\theta_l,\phi_l,\mathbf{s}) j_d(\mathbf{s}) d\mathbf{s}\right|^2 \\
    \text{s.t.} &~~ \int_{\mathcal{S}_\mathrm{T}} \left|j_d(\mathbf{s})\right|^2 d\mathbf{s}= P_t.
\end{align}    
\end{subequations}
The problem (\ref{ref_design_origin}) is a non-convex functional programming, where a Fourier-based approach is developed to address it as follows. The key idea of the Fourier-based approach is to transform the original functional programming in continuous domain into the design of discrete Fourier coefficients by approximating the continuous current waveform using its Fourier series representation. According to the Fourier series lemma in \cite{linglong_CAPA}, the continuous function $j_d(\mathbf{s})$ can be equivalently expressed by the weighted summation of infinite orthogonal Fourier base functions, i.e., 
\begin{align}
    j_d(\mathbf{s})=\sum_{\mathbf{m}=-\infty}^{\infty} w_{\mathbf{m}}\psi_\mathbf{m}(\mathbf{s}),
\end{align}
where $\mathbf{m}=[m_x, m_y]^\mathrm{T}$ and $\sum_{\mathbf{m}=-\infty}^{\infty}$ is defined as $\sum_{m_x=-\infty}^{\infty} \sum_{m_y=-\infty}^{\infty}$. 
Here,  $w_{\mathbf{m}}$ and $\psi_\mathbf{m}(\mathbf{s})$ represent the Fourier coefficients and the orthonormal Fourier base functions, respectively, which are given by
\begin{subequations}
  \begin{align}
    \psi_\mathbf{m}(\mathbf{s})&=\frac{1}{\sqrt{A_\mathrm{T}}}e^{\jmath2\pi\left(\frac{m_x}{L_x} s_x + \frac{m_y}{L_y} s_y \right)},\\
    w_{\mathbf{m}} &=\frac{1}{\sqrt{A_\mathrm{T}}} \int_{\mathcal{S}_\mathrm{T}} j_d(\mathbf{s}) \psi^*_\mathbf{m}(\mathbf{s})d\mathbf{s}.
\end{align}  
\end{subequations}
However, directly dealing with the infinite summation of the Fourier series is challenging in practice.
Considering that the wavenumber-domain spectrum is band-limited, the series are truncated to a finite number of terms.
Accordingly, $j_d(\mathbf{s})$ can be approximated as
\begin{align} \label{approx}
    j_d(\mathbf{s})\approx \sum_{\mathbf{m}=-\mathbf{M}}^{\mathbf{M}} w_{\mathbf{m}}\psi_\mathbf{m}(\mathbf{s}),
\end{align}
where $\mathbf{M}=\left[M_x,M_y\right]^\mathrm{T}$ is the truncation length. As suggested in \cite{linglong_CAPA}, $M_x$ and $M_y$ are set to 
\begin{align}
    M_x=\left\lceil \frac{L_x}{\lambda} \right\rceil,\quad M_y=\left\lceil \frac{L_y}{\lambda} \right\rceil,
\end{align} 
which are the minimum positive integers involving the low-wavenumber and high-power components. Based on (\ref{approx}), the integral terms w.r.t $j_d(\mathbf{s})$ can be transformed into vector forms, expressed as
\begin{subequations} \label{jd_approx}
    \begin{align}
        &\int_{\mathcal{S}_\mathrm{T}} \left|j_d(\mathbf{s})\right|^2 d\mathbf{s} \approx \sum_{\mathbf{m}=-\mathbf{M}}^{\mathbf{M}} |w_{\mathbf{m}}|^2 = \|\mathbf{w}\|^2,\\
        &\int_{\mathcal{S}_\mathrm{T}}  a(\theta_l,\phi_l,\mathbf{s}) j_d(\mathbf{s}) d\mathbf{s} \approx \sum_{\mathbf{m}=-\mathbf{M}}^{\mathbf{M}}  \widetilde{a}_{\mathbf{m}}(\theta_l,\phi_l)w_{
        \mathbf{m}}\notag\\
        &=\mathbf{\widetilde{a}}^\mathrm{T} (\theta_l,\phi_l) \mathbf{w},
    \end{align}
\end{subequations}
where $\widetilde{a}_{\mathbf{m}}(\theta_l,\phi_l)=\int_{\mathcal{S}_T} a(\theta_l,\phi_l,\mathbf{s})\psi_{\mathbf{m}}(\mathbf{s})d\mathbf{s}$. 
Here, $\mathbf{w}$ and $\mathbf{\widetilde{a}}(\theta_l,\phi_l) $ are the vectors with dimensions of $M_{\mathrm{F}}=(2M_x+1)(2M_y+1)$, which collect  all $w_{\mathbf{m}}$ and $\widetilde{a}_{\mathbf{m}}(\theta_l,\phi_l)$.

Accordingly, the problem in (\ref{ref_design_origin}) can be transformed into a conventional finite-dimensional optimization problem, which is written as
\begin{subequations} \label{ref_design_discrete}
\begin{align}
     \max_{\mathbf{w}} &~~ \min_{\theta_l,\phi_l \in \mathrm{\Theta}} ~~ \xi^2(\theta_l,\phi_l) \left|\mathbf{\widetilde{a}}^\mathrm{T} (\theta_l,\phi_l) \mathbf{w}\right|^2\\
    \text{s.t.} &~~ \|\mathbf{w}\|^2= P_t.
\end{align}    
\end{subequations}
However, the problem (\ref{ref_design_discrete}) still cannot be solved directly due to the intractable max-min problem formulation. To address this issue, an auxiliary variable $\beta$ is introduced to equivalently transform the problem, which gives
\begin{subequations} \label{ref_design_relax}
\begin{align}
     \max_{\mathbf{w},\beta} &~~ \beta\\
    \text{s.t.} &~~ 
    \xi^2(\theta_l,\phi_l) \left|\mathbf{\widetilde{a}}^\mathrm{T} (\theta_l,\phi_l) \mathbf{w}\right|^2\geq \beta, \forall \theta_l,\phi_l \in \mathrm{\Theta}, \\
    &~~ \|\mathbf{w}\|^2= P_t. 
\end{align}    
\end{subequations}
The problem (\ref{ref_design_relax}) remains non-convex due to the non-convex constraint (\ref{ref_design_relax}{b}). While it cannot be solved directly, many well-established techniques can be employed to address such a problem, such as semi-definite relaxing (SDR) \cite{SDR}, and successive convex approximation (SCA) \cite{SOCP}. Given the obtained $\mathbf{w}$, the corresponding $j_d(\mathbf{s})$ can be calculated via (\ref{approx}).

\section{ISAC Waveform Design} \label{algorithm}
In this section, a CoV-based algorithm is developed to solve problem (\ref{ISAC_formulation}) based on the reference sensing waveform obtained in Section \ref{sec:EM_ref}. Specifically,   (\ref{ISAC_formulation}) is first converted to an unconstrained problem via Lagrangian transformation. Then, we derive the optimal ISAC waveform structure based on CoV. Besides, the optimal Lagrangian multiplier is searched via the Bisection method. Furthermore,  the computational complexity of the proposed algorithm is discussed.

\subsection{CoV-Based Algorithm}
To better observe the objective function in (\ref{ISAC_formulation}), we first expand the squared terms and drop the constant terms, which yields
\begin{align}\label{simplify_obj}
    \widetilde{f}(j(\mathbf{s}))=& \rho \sum_{k=1}^K\int_{\mathcal{S}_\mathrm{T}} \int_{\mathcal{S}_\mathrm{T}} j^*(\mathbf{s}) H^*_k(\mathbf{s})  H_k(\mathbf{s}')  j(\mathbf{s}') d\mathbf{s}' d\mathbf{s}  \notag\\
    & - \sum_{k=1}^K 2\rho\mathfrak{R}\left \{c_k \int_{\mathcal{S}_\mathrm{T}} H_k^*(\mathbf{s})  j^*(\mathbf{s})  d\mathbf{s} \right \} \notag \\
    &-  (1-\rho)\int_{\mathcal{S}_\mathrm{T}}  2\mathfrak{R} \left \{j^*(\mathbf{s}) j_{d}(\mathbf{s}) \right\} d\mathbf{s} .
\end{align}
Here, $|c_k|^2$ and $\int_{\mathcal{S}_\mathrm{T}} j^*d(\mathbf{s}) j_d(\mathbf{s}) d\mathbf{s}$ are independent of $j(\mathbf{s})$, thus they can be treated as constants. Moreover, when the constraint (\ref{ISAC_formulation}{b}) is satisfied, the term $\int_{\mathcal{S}_\mathrm{T}} |j(\mathbf{s}) |^2 d\mathbf{s}$ can also be regarded as a constant and omitted from the objective function. As a result, the problem (\ref{ISAC_formulation}) can be simplified as
\begin{align} \label{ISAC_formulation2}
    \min_{j(\mathbf{s})} &~~ \widetilde{f}(j(\mathbf{s}))  \quad\mathrm{s.t.}  \int_{\mathcal{S}_\mathrm{T}} \left|j(\mathbf{s})\right|^2 d\mathbf{s}= P_t.
\end{align} 
According to \cite{LiuFan_OptimalWaveform}, the problem (\ref{ISAC_formulation2}) is a convex functional programming problem with an equality constraint, where strong duality holds.

To solve the above problem, we first transform it into an unconstrained optimization problem via the Lagrangian transformation. By introducing a Lagrangian multiplier $\mu$  associated with the constraint, the corresponding Lagrangian function for (\ref{ISAC_formulation2}) can be expressed as
\begin{align}
    \mathcal{L}(j(\mathbf{s}),\mu)=& \widetilde{f}(j(\mathbf{s})) + \mu \left( \int_{\mathcal{S}_\mathrm{T}} |j(\mathbf{s}) |^2 d\mathbf{s}-P_t\right). 
\end{align}
Given an optimal $\mu^\star$, the optimal $j(\mathbf{s})$ can be obtained by minimizing the Lagrangian function $\mathcal{L}(j(\mathbf{s}),\mu^\star)$, which is an unconstrained functional programming. In the following, we employ the CoV to derive the optimal waveform structure and search the corresponding Lagrangian multiplier, with which the optimal ISAC waveform can be obtained. 

Prior to deriving the optimal waveform structure via CoV, we introduce a fundamental lemma as follows.
\begin{lemma} \label{lem_cov}
   (Fundamental Lemma of CoV):  Consider an arbitrary smooth function $f_0(\mathbf{s})$, defined on an open set $\mathcal{S}$ in the complex space, which falls into $0$ on the boundary $\partial\mathcal{S}$, i.e.,
\begin{align}
      f_0(\mathbf{s})=0, \forall\mathbf{s} \in \partial\mathcal{S}.
\end{align}
Suppose there exists a continuous function  $f_1(\mathbf{s})$ on $\mathcal{S}$ satisfying
\begin{align} \label{lemma_eq2}
    \Re\left\{\int_{\mathcal{S}} f_0^*(\mathbf{s}) f_1(\mathbf{s}) d\mathbf{s} \right\} =0.
\end{align}
  Then, it must hold that
\begin{align}
    f_1(\mathbf{s})=0, \forall\mathbf{s}  \in\mathcal{S}.
\end{align}
\end{lemma}
\begin{IEEEproof}
    Assume that $f_{1}(\mathbf{s})\neq 0$ is defined in $S$ satisfying (\ref{lemma_eq2}). In a special case of $f_{0}(\mathbf{s})=g(\mathbf{s})f_{1}(\mathbf{s})$ where $g(\mathbf{s})>0,\forall \mathbf{s} \in \mathcal{S}$ and 
$g(\mathbf{s})=0,\forall\mathbf{s} \in\partial S$, the condition  (\ref{lemma_eq2}) becomes
$\Re\{\int_\mathcal{S} g(\mathbf{s}) | f_{1}(\mathbf{s})|^{2}d\mathbf{s}\}=0$, which indicates $f_{1}(\mathbf{s})= 0$. This contradicts with the prior assumption. The proof ends.
\end{IEEEproof}
Based on \textbf{Lemma \ref{lem_cov}}, given  $\mu^\star$, the optimal waveform structure can be derived from the necessary condition for the optimal $j(\mathbf{s})$ that minimizes $\mathcal{L}(j(\mathbf{s}),\mu^\star)$. The result is given in the following theorem.

\begin{theorem} \label{theo_optimal_structure}
    (Optimal Structure of ISAC Waveform): When given an optimal $\mu^\star$, the waveform structure that minimize the functional $\mathcal{L}(j(\mathbf{s}),\mu^\star)$ can be given by
    \begin{align} \label{optimal_structure}
    \mu^\star j(\mathbf{s}) &= -\rho \sum_{k=1}^K H^*_k(\mathbf{s})\int_{\mathcal{S}_\mathrm{T}}   H_k(\mathbf{s}')  j(\mathbf{s}') d\mathbf{s}'  \notag \\
    & +  \rho \sum_{k=1}^K c_k H^*_k(\mathbf{s})  + (1-\rho) j_{d}(\mathbf{s}).
\end{align}
\end{theorem}
\begin{IEEEproof}
    Please refer to Appendix \ref{app_a}.
\end{IEEEproof}

 According to \textbf{Theorem \ref{theo_optimal_structure}}, the optimal $j(\mathbf{s})$ satisfies (\ref{optimal_structure}), which is a Fredholm integral
equation of the second kind. In the following, we will solve (\ref{optimal_structure}) and obtain the optimal $\mu^\star$. First, (\ref{optimal_structure}) can be re-written as
\begin{align} \label{optimal_structure2}
  \! \!\! \mu^\star  j(\mathbf{s}) \!\!=\! -\rho \sum_{k=1}^K H^*_k(\mathbf{s}) z_k   \!+\!  \rho \sum_{k=1}^K c_k H^*_k(\mathbf{s})  \!+\! (1\!-\!\rho) j_{d}(\mathbf{s}),
\end{align}
where we denote
\begin{align} \label{z_expression}
    z_k = \int_{\mathcal{S}_\mathrm{T}}   H_k(\mathbf{s})  j(\mathbf{s}) d\mathbf{s}.
\end{align}
Then, we multiply both sides of (\ref{optimal_structure2}) by 
$H_i(\mathbf{s})$ and integrate over 
$\mathbf{s}$, which yields
\begin{align} \label{optimal_structure3}
    \mu^\star \int_{\mathcal{S}_\mathrm{T}} H_i(\mathbf{s}) j(\mathbf{s}) d\mathbf{s} =& -\rho \sum_{k=1}^K z_k \int_{\mathcal{S}_\mathrm{T}} H_i(\mathbf{s}) H^*_k(\mathbf{s}) d\mathbf{s} \notag  \\
    &+  \rho \sum_{k=1}^K c_k \int_{\mathcal{S}_\mathrm{T}} H_i(\mathbf{s}) H^*_k(\mathbf{s}) d\mathbf{s} \notag \\
    &+ (1-\rho) \int_{\mathcal{S}_\mathrm{T}} H_i(\mathbf{s}) j_{d}(\mathbf{s}) d\mathbf{s}.
\end{align}
On the left-hand side, it can be seen that the expression is exactly in the consistent form of (\ref{z_expression}).
Therefore, the equation in (\ref{optimal_structure3}) can be further rewritten as
\begin{align}
    \mu^\star z_i = -\rho \sum_{k=1}^K z_k q_{i,k} +  \rho \sum_{k=1}^K c_k q_{i,k} + (1-\rho) u_i, 
\end{align}
where we denote
\begin{subequations}\label{q_u_definition}
    \begin{align} 
    q_{i,k} &= \int_{\mathcal{S}_\mathrm{T}} H_i(\mathbf{s}) H^*_k(\mathbf{s}) d\mathbf{s}, \forall i,k=1,\cdots,K\\
    u_i &=\int_{\mathcal{S}_\mathrm{T}} H_i(\mathbf{s}) j_{d}(\mathbf{s}) d\mathbf{s}, ~~\forall i=1,\cdots,K.
\end{align}
\end{subequations}
Here, $q_{i,k}$ characterizes the channel correlation among users, while $u_i$ captures the correlations between the sensing reference signal and the communication channel. By stacking all the $z_i$ into a vector $\mathbf{z}$, (\ref{optimal_structure3}) can be compactly  written  in a matrix form, i.e.,
\begin{align} \label{optimal_structure_mat}
    \mu^\star \mathbf{z} = -\rho \mathbf{Q} \mathbf{z} +  \rho  \mathbf{Q} \mathbf{c}  + (1-\rho) \mathbf{u},
\end{align}
where
\begin{subequations}
   \begin{align}
    \mathbf{z} &= \begin{bmatrix}
        z_1 &\cdots & z_K
    \end{bmatrix}^\mathrm{T},\\
    \mathbf{c} &= \begin{bmatrix}
        c_1 &\cdots & c_K
    \end{bmatrix}^\mathrm{T},\\
    \mathbf{u} &= \begin{bmatrix}
        u_1 &\cdots & u_K
    \end{bmatrix}^\mathrm{T},\\
    \mathbf{Q} &= \begin{bmatrix}
        q_{1,1} &\cdots & q_{1,K}\\
         &\ddots & \\
        q_{K,1} &\cdots & q_{K,K}\\
    \end{bmatrix}.
\end{align} 
\end{subequations}

Since $j(\mathbf{s})$ appears only in $\mathbf{z}$, we separate $\mathbf{z}$ from other components, which can be expressed as
\begin{align}\label{z_expression_mat}
    \mathbf{z} & =  (\mu^\star \mathbf{I}+\rho \mathbf{Q} )^{-1}(\rho  \mathbf{Q} \mathbf{c}  + (1-\rho) \mathbf{u}).
\end{align}
On the right-hand side of (\ref{z_expression_mat}),  $\mu^\star$ is the only unknown. To obtain $\mu^\star$, an equation w.r.t. $\mu^\star$ can be constructed based on the  constraint (\ref{ISAC_formulation}{b}), which is given in the following theorem. 
\begin{theorem} \label{theo_lambda_eq}
  (Condition of $\mu^\star$) : The optimal $\mu^\star$ that guarantees the power constraint and minimizes the objective function can be obtained by solving the following equation:
    \begin{align} \label{lambda_eq}
    &\rho^2   \mathbf{z}^\mathrm{H} (\mu) \mathbf{Q} \mathbf{z}(\mu)  \! -\!\!2 \rho^2 \Re\left\{ \mathbf{c}^\mathrm{H} \mathbf{Q} \mathbf{z} (\mu) \right\}  \notag \\
    &\!-\!\!2 \rho (1\!-\!\rho) \Re\left\{ \mathbf{z}(\mu)^\mathrm{H} \mathbf{u}   \right\} \!+\! \widetilde{c} \!= \! \mu^2 P_t,
\end{align}
where the function $\mathbf{z}(\mu)$ is indicated in (\ref{z_expression_mat}), and $\widetilde{c}= \rho^2 \mathbf{c}^\mathrm{H} \mathbf{Q} \mathbf{c} + (1-\rho)^2 P_t + 2 \rho (1-\rho) \Re\left\{\mathbf{c}^\mathrm{H} \mathbf{u}  \right\}$ combines all the terms that are unrelated to $\mathbf{z}$ and $\mu$. 
\end{theorem}
\begin{IEEEproof}
    Please refer to  Appendix \ref{app_b}.
\end{IEEEproof}
While directly solving (\ref{lambda_eq}) is challenging, 
 $\mu^\star$ can be numerically
calculated using the Bisection method.  After obtaining the optimal $\mu^\star$,  it can be substituted into (\ref{z_expression_mat}) and obtain all the $z_i$. Subsequently, with the obtained $z_i$ and $\mu^\star$, the optimal waveform $j(\mathbf{s})$ can be calculated by
\begin{align}\label{final_j}
       \!\!\!\!j(\mathbf{s}) \!=\! -\frac{\rho}{\mu^\star} \!\sum_{k=1}^K\! H^*_k(\mathbf{s}) z_k   \!+ \! \frac{\rho}{\mu^\star} \!\sum_{k=1}^K \!c_k H^*_k(\mathbf{s})  \!+\! \frac{(1\!-\!\rho)}{\mu^\star} j_{d}(\mathbf{s}).
\end{align}
It is also worth mentioning that, for computational simplicity, the objective function in   (\ref{ISAC_formulation}{a}) can be calculated using the obtained $\mu^\star$ and matrix calculation, thereby avoiding the need to compute the complicated integrals directly.  From the simplified form of  (\ref{ISAC_formulation}{a}) in (\ref{simplify_obj}), the first and second terms contains $\int_{\mathcal{S}_\mathrm{T}}   H_k(\mathbf{s})  j(\mathbf{s}) d\mathbf{s}$, which corresponds to $z_k$. Consequently, the first two terms in (\ref{simplify_obj}) can be calculated by
\begin{align}
    &\rho \sum_{k=1}^K z_k^* z_k   - \sum_{k=1}^K 2\rho\mathfrak{R}\left \{c_k z_k^* \right \} =\rho \| \mathbf{z}\|^2 -2\rho\mathfrak{R}\left \{ \mathbf{z}^\mathrm{H} \mathbf{c} \right \}.
\end{align}
Similarly, the third term in (\ref{simplify_obj}) can be computed using $\mathbf{u}$, $\mathbf{c}$ and $\mathbf{z}$. Considering the optimal structure in (\ref{optimal_structure2}), we multiple both sides by $j^*_d(\mathbf{s})$ and integrate over $\mathbf{s}$, which yields
\begin{align} \label{optimal_structure5}
    \mu^\star \int_{\mathcal{S}_\mathrm{T}} j^*_d(\mathbf{s}) j(\mathbf{s}) d\mathbf{s} =& -\rho \sum_{k=1}^K z_k \int_{\mathcal{S}_\mathrm{T}} j^*_d(\mathbf{s}) H^*_k(\mathbf{s}) d\mathbf{s} \notag  \\
    &+  \rho \sum_{k=1}^K c_k \int_{\mathcal{S}_\mathrm{T}} j^*_d(\mathbf{s}) H^*_k(\mathbf{s}) d\mathbf{s} \notag \\
    &+ (1-\rho) \int_{\mathcal{S}_\mathrm{T}} j^*_d(\mathbf{s}) j_{d}(\mathbf{s}) d\mathbf{s}.
\end{align}
It can be seen that the right-hand side of  (\ref{optimal_structure5}) involves  $\int_{\mathcal{S}_\mathrm{T}}  H_k(\mathbf{s}) j_d(\mathbf{s}) d\mathbf{s}$, which corresponds to $u_k$ in (\ref{q_u_definition}{b}). Meanwhile, the reference waveform $j_d(\mathbf{s})$ has a total power of $P_t$ as described in Section \ref{subsec:reference_signal}, i.e., $\int_{\mathcal{S}_\mathrm{T}} j^*_d(\mathbf{s}) j_{d}(\mathbf{s}) d\mathbf{s} = P_t$. Accordingly, the third term in (\ref{simplify_obj}) can be expressed as
\begin{align}
     \int_{\mathcal{S}_\mathrm{T}} j^*_d(\mathbf{s}) j(\mathbf{s}) d\mathbf{s} =& -\frac{\rho}{\mu^\star} \sum_{k=1}^K z_k u_k^*  +  \frac{\rho}{\mu^\star} \sum_{k=1}^K c_k u_k^*  + \frac{1-\rho}{\mu^\star} P_t \notag \\
     =& -\frac{\rho}{\mu^\star} \mathbf{u}^\mathrm{H} \mathbf{z}+  \frac{\rho}{\mu^\star} \mathbf{u}^\mathrm{H} \mathbf{c}  + \frac{1-\rho}{\mu^\star} P_t.
\end{align}
Therefore, (\ref{simplify_obj}) can be calculated by
\begin{align} \label{obj_value_mat}
     \widetilde{f}(j(\mathbf{s}))=&\rho \| \mathbf{z}\|^2 -2\rho\mathfrak{R}\left \{ \mathbf{z}^\mathrm{H} \mathbf{c} \right \} \notag \\
    &\!\!\!\!-2(1\!-\!\rho) \mathfrak{R}\left \{ -\frac{\rho}{\mu^\star} \mathbf{u}^\mathrm{H} \mathbf{z}\!+\!  \frac{\rho}{\mu^\star} \mathbf{u}^\mathrm{H} \mathbf{c}  \!+ \!\frac{1\!-\!\rho}{\mu^\star} P_t\right\}.
\end{align}
Additionally, the constant terms omitted from (\ref{ISAC_formulation}{a}) can be directly obtained, namely, $\|\mathbf{c}\|^2$, $\int_{\mathcal{S}_\mathrm{T}} \left|j(\mathbf{s})\right|^2 d\mathbf{s}= P_t$ and $\int_{\mathcal{S}_\mathrm{T}} \left|j_d(\mathbf{s})\right|^2 d\mathbf{s}= P_t$.
The overall algorithm is summarized in \textbf{Algorithm 1}.

\begin{algorithm}[!t] 
\caption{CoV-based Algorithm for Joint  Communication MUI and Sensing Beampattern Mismatch Minimization 
Problem in CAPA-aided ISAC System} 
    \begin{algorithmic}[1] 
        \REQUIRE ~  CAPA system parameters, communication channel function, target directions, communication symbols, and transmit power;
        \STATE Generate reference sensing waveform  $j_d(\mathbf{s})$ by solving (\ref{ref_design_relax}) and calculate $\mathbf{u}$ with (\ref{q_u_definition}{b});
        \STATE Compute the CAPA channel correlation matrix $\mathbf{Q}$ in (\ref{q_u_definition}{a});
        \STATE Utilize the Bisection search to solve the equation (\ref{lambda_eq}) and find the optimal Lagrangian multiplier $\mu^\star$;
        \STATE Obtain the optimal $j(\mathbf{s})$ using (\ref{final_j}) and the minimum objective function value using (\ref{obj_value_mat}).
    \end{algorithmic}
\end{algorithm}

\subsection{Integral Calculations and Algorithm Complexity}
In the process of obtaining the reference waveform and calculating $\mathbf{u}$ and $\mathbf{Q}$ in \textbf{Algorithm 1}, several integrals need to be computed, where the Gauss-Legendre quadrature is employed. Specifically, Gauss-Legendre quadrature approximates an integral using weighted sums of function values at specific points, which facilitates the computation of the integrals. For a function that integrates over $[a, b]$, the approximation can be expressed as follows:
\begin{align}
    \int_{a}^{b} f(x) dx \approx \frac{b-a}{2}\sum_{n=1}^{N} \nu_n f\left(\frac{b-a}{2}\zeta_n+\frac{a+b}{2}\right),
\end{align}
where $N$ represents the number of sample points for integrals, $\nu_n$ denotes the corresponding quadrature weights and $\zeta_n$ are the roots of the $n$-th Legendre polynomial. As $N$ increases, the integral approximation becomes more accurate. For demonstration, we take the computation of $\mathbf{u}$ in (\ref{q_u_definition}{b}) as an example: 
    \begin{align} 
    u_i &=\int_{\mathcal{S}_\mathrm{T}} H_i(\mathbf{s}) j_{d}(\mathbf{s}) d\mathbf{s} \notag\\
    &=\int_{-\frac{L_y}{2}}^{\frac{L_y}{2}} \int_{-\frac{L_x}{2}}^{\frac{L_x}{2}} H_i(s_x,s_y) j_{d}(s_x,s_y) ds_x ds_y \notag \\
    &\approx \frac{L_xL_y}{4} \sum_{n_x=1}^{N}\sum_{n_y=1}^{N} \mu_{n_x}\mu_{n_y} H_i\left(\frac{L_x}{2}\zeta_{m_x},\frac{L_y}{2}\zeta_{m_y}\right) \notag \\
    &~~~~~~~~~~~~~~~~~~~~~~~~  \times j_{d}\left(\frac{L_x}{2}\zeta_{m_x},\frac{L_y}{2}\zeta_{m_y}\right).
\end{align}
The rest integrals are computed in the same manner as the above calculation.

In addition, the computational complexity of the algorithm is analyzed as follows. 
\begin{itemize}
    \item In generating the reference waveform, the complexity stems from the calculations of $\mathbf{\widetilde{a}}(\theta_l,\phi_l) $ in (\ref{jd_approx}{b}) and solving the problem of (\ref{ref_design_relax}). To obtain $\mathbf{\widetilde{a}}(\theta_l,\phi_l)$, $M_{\mathrm{F}}$ Gauss-Legendre quadratures need to be calculated, which have a total computational complexity of $\mathcal{O}(M_{\mathrm{F}}  N^2)$. For  solving the problem (\ref{ref_design_relax}), the SDR has a computational complexity of $\mathcal{O}(M_{\mathrm{F}}^{4.5})$, while SCA incurs an approximate computational complexity of $\mathcal{O}(I_0 M_{\mathrm{F}}^3)$, where $I_0$ represents iteration number of SCA. 

    \item In obtaining $j(\mathbf{s})$ of (\ref{final_j}), the computation of $\{\mathbf{Q},\mathbf{u}\}$, the optimal $\mu^\star$, and $\mathbf{z}$ contributes to the complexity.  Computing $\mathbf{Q}$ and $\mathbf{u}$ involves $K^2$ and $K$ integrals, which have computational complexities of  $\mathcal{O}(K^2 N^2)$ and $\mathcal{O}(K N^2)$, respectively. To find the optimal $\mu^\star$, (\ref{lambda_eq}) is solved with the Bisection search. It incurs a computational complexity of $\mathcal{O}(\log_2(1/\epsilon_0))$, where $\epsilon_0$ denotes the convergence parameter of the Bisection search method. Finally, the calculation of $\mathbf{z}$ in (\ref{z_expression_mat}) involves matrix inversion and multiplications.  The inverse operation has a complexity of $\mathcal{O}(K^3)$, while the matrix operations incur a computational complexity of $\mathcal{O}(K^2)$.
    
\end{itemize}
 In summary, the total computational complexity of the \textbf{Algorithm 1} can be approximated as $\mathcal{O}(M_{\mathrm{F}}^{4.5})$, neglecting the lower-order terms.

\section{Simulation Results} \label{simulation}

  \begin{figure}[!t]
    \centering
    \includegraphics[scale=0.46]{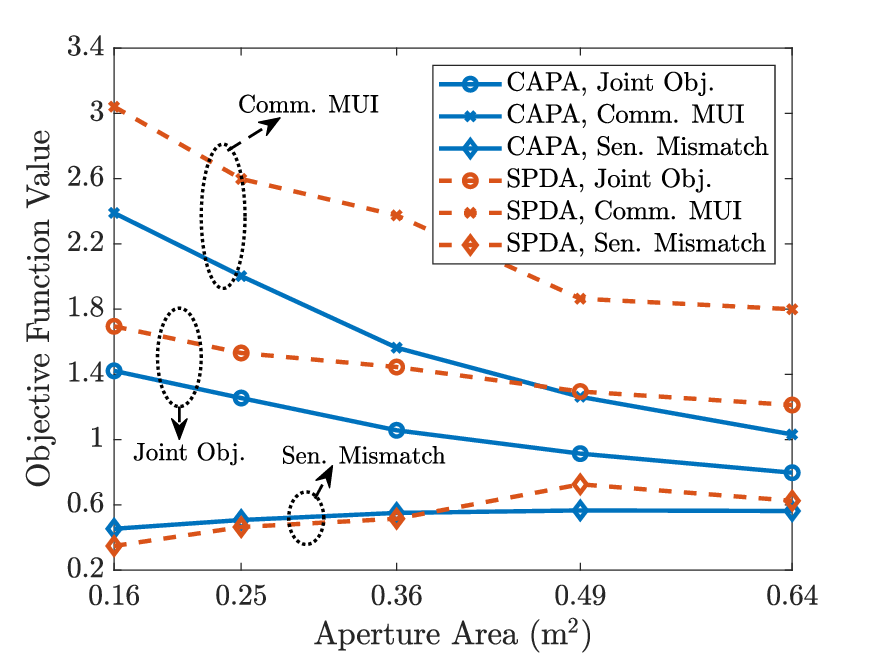}
    \caption{The impact of the CAPA aperture size on the overall system performance, sensing beampattern mismatch, and communication MUI.}
    \label{fig:Aper}
\end{figure}
In this section, simulation results are presented to validate the effectiveness of the proposed system and algorithm. Unless otherwise specified, the default parameters are configured as follows.  The carrier frequency center is $2.4$ GHz, corresponding to a wavelength of $\lambda=0.125$ m.  The free-space impedance is set to $\eta=120\pi$ $\Omega$. For the CAPA transmitter, it is centered at the origin, with length and width of $L_x=L_y=0.6$ m, resulting in a total aperture area of $A_\mathrm{T}=0.36$ m$^2$. The transmit power is fixed at $P_t=5$ A$^2$. For the sensing task, $T=3$ targets are considered, which are located in the directions of ($45^\circ$, $15^\circ$), ($-60^\circ$, $45^\circ$) and ($30^\circ$, $60^\circ$), respectively.  For the communication scenarios, $K=4$ communication users are served, where they are randomly distributed within a circular area centered at ($20$, $-20$, $30$)  m with a radius of  $10$ m. Besides, quadrature phase shift keying (QPSK) is adopted for modulation. We define the transmit signal-to-noise ratio (SNR) for user $k$ as SNR$_k$=$P_t/\sigma_k^2$, and all users have the same SNR of $10$ dB.  The tradeoff coefficient between sensing and communication is set to $\rho=0.5$.  For the calculations of the integrals, the Gauss–Legendre quadrature employs $N = 20$ sampling points. All simulation results are averaged over $1000$ Monte Carlo trials to ensure statistical reliability.

To evaluate the performance of the proposed design, the conventional SPDA is adopted as a benchmark, whose optimization formulation and solution method are provided in \cite{LiuFan_OptimalWaveform}. For a fair comparison, both the sensing and communication channel responses are modeled consistently with the CAPA framework, as detailed in \cite{Zhaolin_CAPA_Optimal_BF}. Moreover, the proposed ISAC system is compared with two baseline cases: a CAPA-aided sensing-only system (Sen.-Only) and a CAPA-aided communication-only system (Comm.-Only), corresponding to $\rho=0$ and $\rho=1$, respectively.

 \begin{figure}[!t]
    \centering
    \includegraphics[scale=0.46]{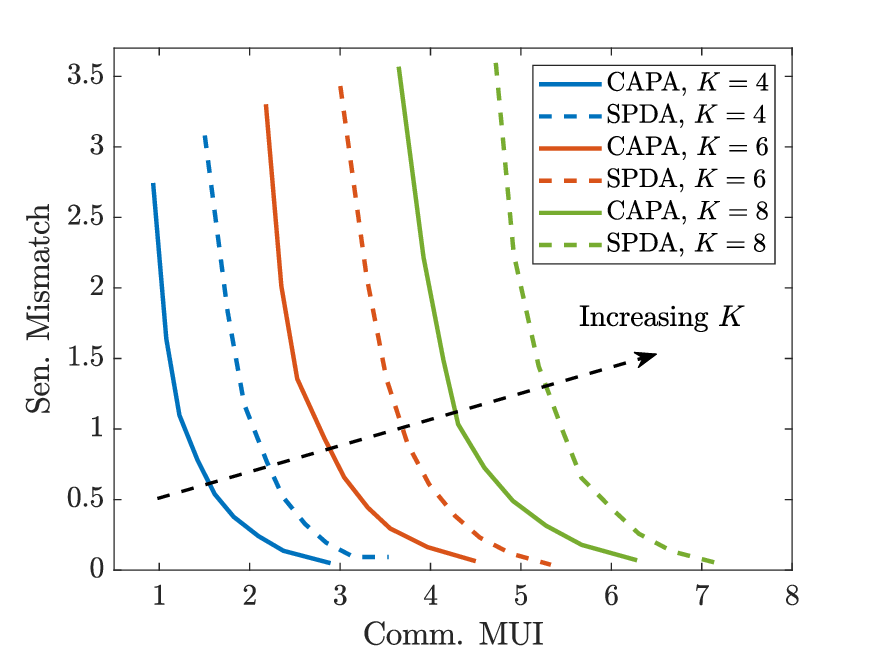}
    \caption{Tradeoff between the communication MUI and the sensing beampattern mismatch under varying $K$ with different arrays.}
    \label{fig:differentK}
\end{figure}

 \begin{figure}[!t]
    \centering
    \includegraphics[scale=0.46]{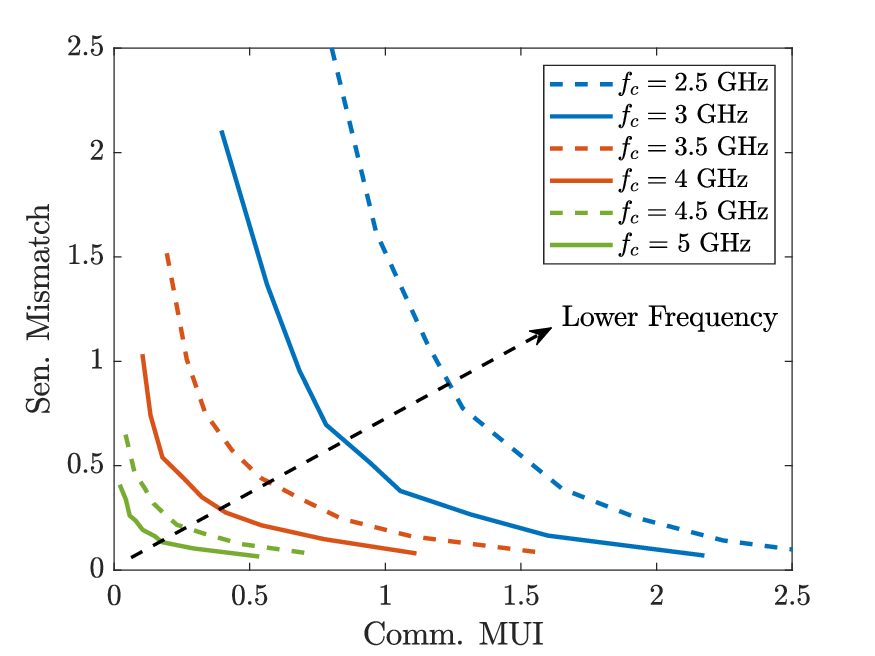}
    \caption{Tradeoff between the communication MUI and the sensing beampattern mismatch under different carrier frequencies.}
    \label{fig:freq}
\end{figure}

\subsection{Overall Performance and Tradeoff}
 We first present the overall performance and the tradeoff relationships between sensing and communication in Fig. \ref{fig:Aper}-\ref{fig:freq}, in terms of the sensing beampattern mismatch (Sen. Mismatch) and communication MUI (Comm. MUI). In Fig. \ref{fig:Aper}, the impact of the CAPA aperture size on system performance is illustrated and compared to the SPDA benchmark, where the aperture size varies from $0.16$ to $0.64$ m$^2$. It can be observed that the overall objective function value (Joint Obj.) decreases as the aperture size increases, since a larger aperture provides more DoFs for mitigating MUI and forming sharper beams, thereby improving the overall performance. Furthermore, the Comm. MUI exhibits a sharp reduction with increasing aperture size, while the Sen. Mismatch remains relatively stable. This indicates that the increased DoFs can significantly reduce the interference among users without distorting the shape of the beampattern for sensing. In addition, CAPA consistently outperforms SPDA due to its continuous-aperture nature, which offers substantially high DoFs for joint waveform design.

Fig. \ref{fig:differentK} illustrates the tradeoff behaviors of the proposed system under different user numbers ($K=4,6,8$). Obviously, an inverse relationship can be observed between the Sen. Mismatch and the Comm. MUI, reflecting the intrinsic tradeoff between sensing and communication performance. Moreover, for a fixed Sen. Mismatch level, the Comm. MUI decreases as the number of users reduces, since fewer users result in weaker interference. Conversely, increasing $K$ slightly degrades sensing performance, where more DoFs are devoted to interference suppression and thus sensing alignment is restricted. In addition, consistent with the previous observations, CAPA achieves a more favorable tradeoff than SPDA, where both sensing and communication performance are improved.

 Fig. \ref{fig:freq} demonstrates the impact of carrier frequency  $f_c$  on the sensing–communication tradeoff performance. As with previous results, the proposed CAPA-aided ISAC system yields better performance than the conventional SPDA across all frequency bands. Moreover, it can be observed that increasing the carrier frequency enhances both sensing and communication performance, which is in alignment with the findings in \cite{freq}. Specifically, when the carrier frequency is $f_c=5$ GHz, the Sen. Mismatch and the Comm. MUI are reduced to within 0.5, approximately one-fifth of their values at $2.5$ GHz. This improvement stems from the higher spatial DoFs and finer angular resolution achievable at higher frequencies, enabling more precise beampattern control and interference suppression.

 \subsection{Communication Performance}
  \begin{figure}[!t]
  \centering
    \includegraphics[scale=0.46]{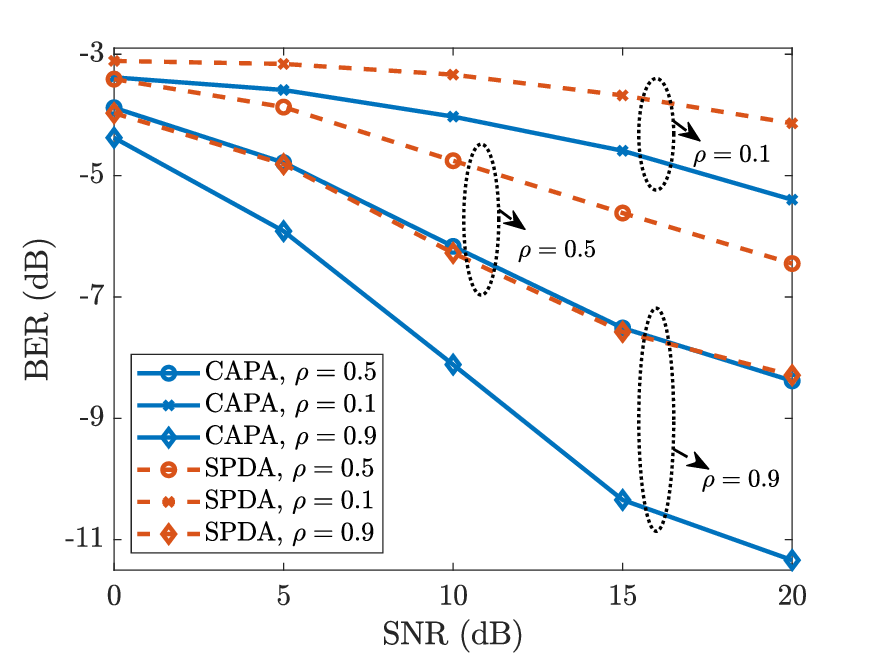}
    \caption{Average BER (dB) versus different transmit SNR (dB) for CAPA under different $\rho$, and comparisons with SPDA.}
    \label{fig:BER1}
\end{figure}
  \begin{figure}[!t]
  \centering
    \includegraphics[scale=0.46]{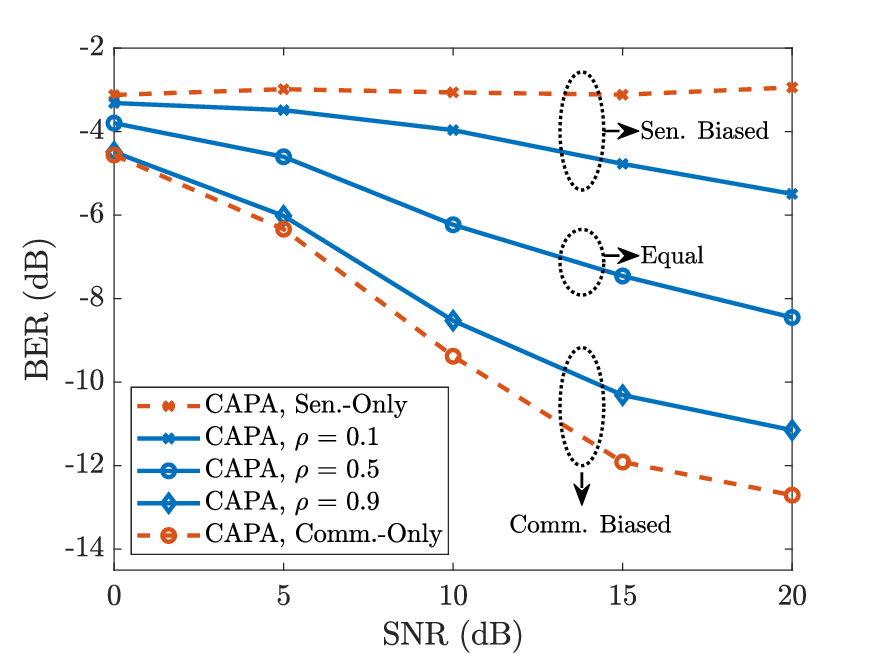}
    \caption{Average BER (dB) versus different transmit SNR (dB): comparisons with the CAPA-aided sensing-only and communication-only cases.}
    \label{fig:BER2}
\end{figure}

 \begin{figure}[!t]
  \centering
    \includegraphics[scale=0.46]{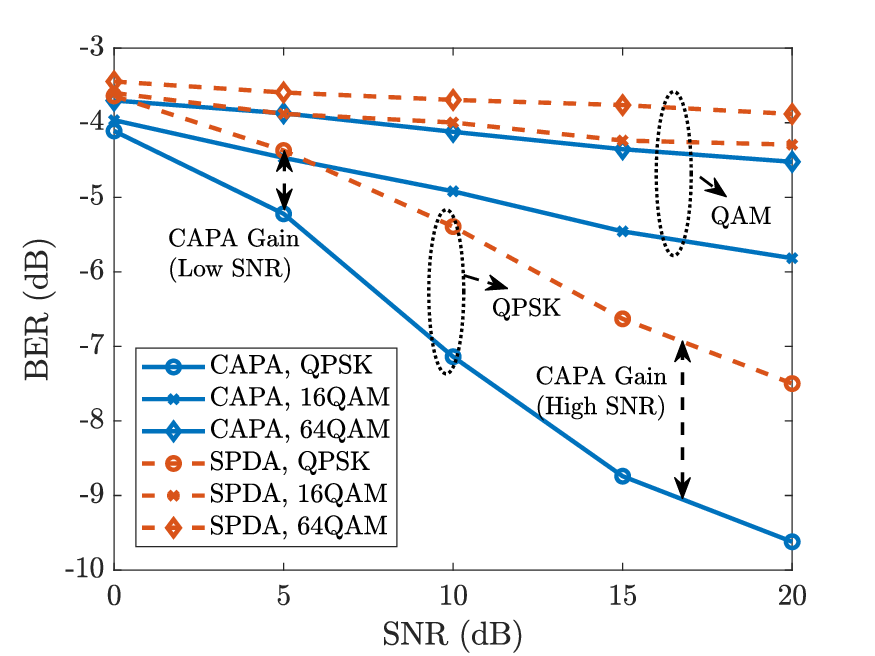}
    \caption{Average BER (dB) versus different transmit SNR (dB) for CAPA under different modulations.}
    \label{fig:BER3}
\end{figure}

In this subsection, we evaluate the communication performance in terms of BER. While transmit SNR is defined as $\mathrm{SNR}=P_t/\sigma_k^2$, the noise power can be calculated with a given transmit power. Accordingly, the BER is obtained based on the received signal model in (\ref{decode_receive_model}) with the corresponding MUI. First, Fig. \ref{fig:BER1} shows the BER results under different weighting coefficients $\rho$ ($\rho=0.1,0.5,0.9$). As expected, the BER decreases with increasing SNR due to the reduction of noise power. Moreover, higher $\rho$ values lead to lower BER, since a larger $\rho$ indicates that the system is biased towards the communication functions, thereby allocating more power to suppress MUI and improving symbol recovery accuracy. In addition, the proposed CAPA-aided system consistently achieves lower BER than the SPDA-aided counterpart across all $\rho$, which further validates the superiority of CAPA in communication performance improvement.

 Similarly, Fig. \ref{fig:BER2} compares the BER of the CAPA-aided ISAC system with the CAPA-aided communication-only and sensing-only systems. As observed, the BER of different $\rho$ values falls within a certain range, where the communication-only and sensing-only cases serve as the lower and upper bounds, respectively. As $\rho$ increases, the system becomes more communication-oriented, and its BER approaches the lower bound. Conversely, a smaller $\rho$ drives the system toward the sensing-dominated regime. In the sensing-only case, the BER remains nearly constant around $-3$ dB regardless of the SNR, since the system fully prioritizes sensing and no MUI suppression is performed. By contrast, when communication is considered only, the BER reaches its minimum because all spatial DoFs are utilized for MUI mitigation. 

Fig. \ref{fig:BER3} illustrates the impact of different modulation schemes on the BER, where ``QPSK'', ``16QAM'' and ``64QAM'' are evaluated. As expected, the BER decreases with increasing SNR. Meanwhile, the BER gaps among different modulations increase as the SNR grows. It is also evident that higher-order modulations yield worse BER performance, since denser constellation points are more sensitive to noise and interference during symbol detection. Furthermore, across all modulation schemes, the CAPA consistently achieves lower BER than the SPDA, with the performance gap being more pronounced for lower-order modulations. For instance, at $\mathrm{SNR}=20$ dB, the CAPA achieves nearly $3$ dB lower BER than SPDA under ``QPSK'', while the gap reduces to about $2$ dB for ``16QAM''.

 \subsection{Sensing Performance}
 \begin{figure*}
	\centering
	\subfigure[]{
		\includegraphics[width=4.2cm]{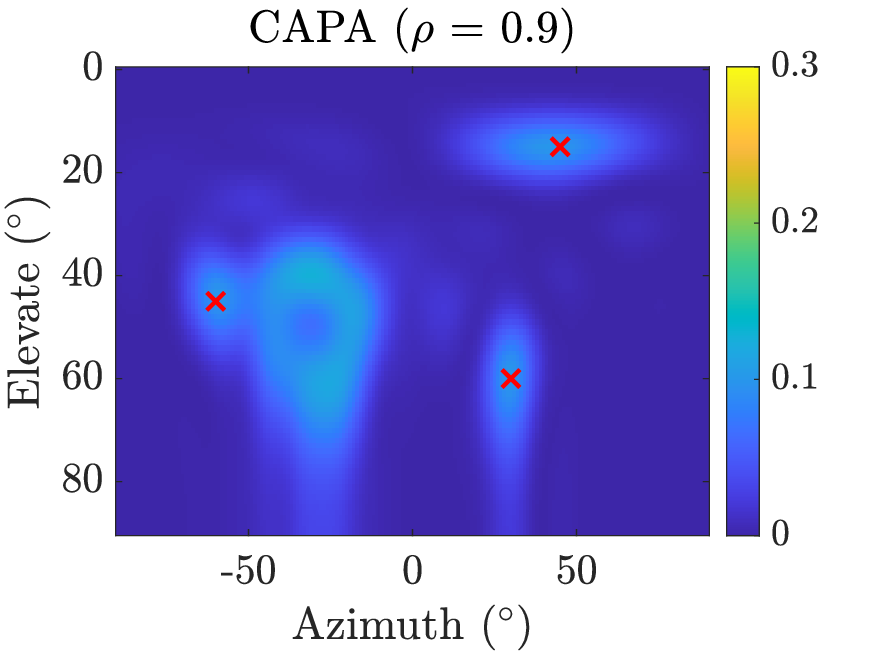}}
	\subfigure[]{
		\includegraphics[width=4.2cm]{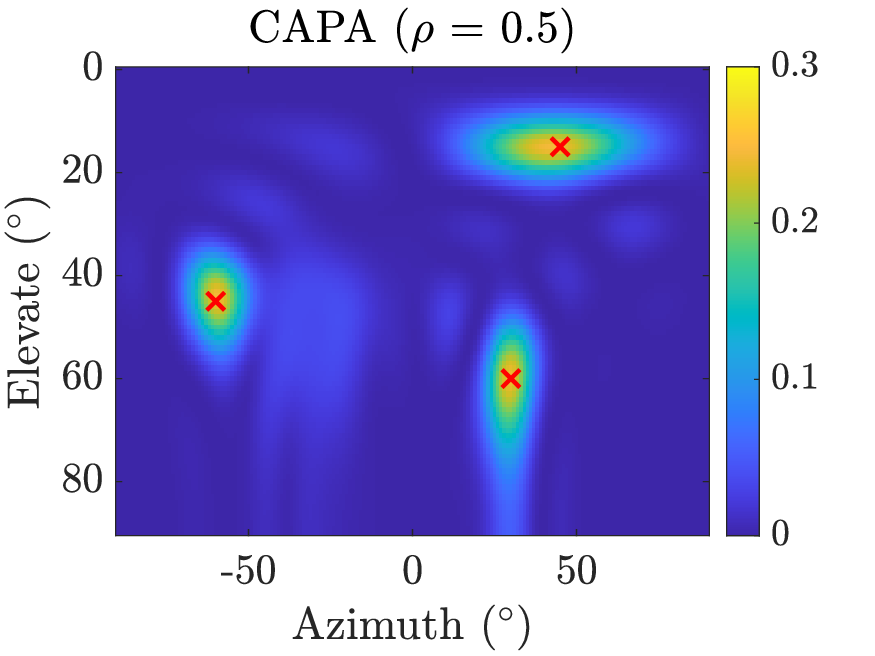}}	
	\centering
	\subfigure[]{
		\includegraphics[width=4.2cm]{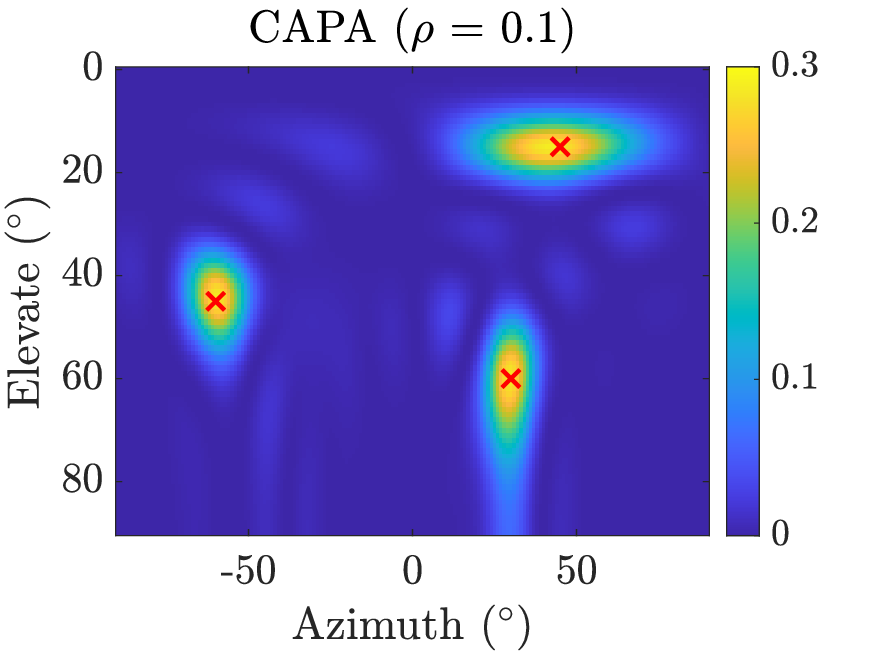}}
	\centering
	\subfigure[]{
		\includegraphics[width=4.2cm]{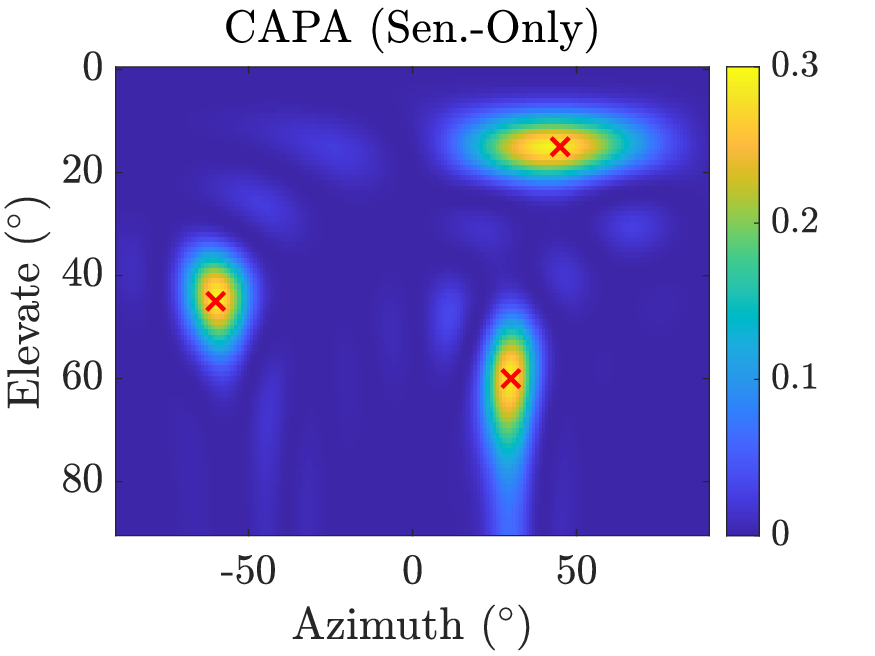}}
        \centering
        \subfigure[]{
		\includegraphics[width=4.2cm]{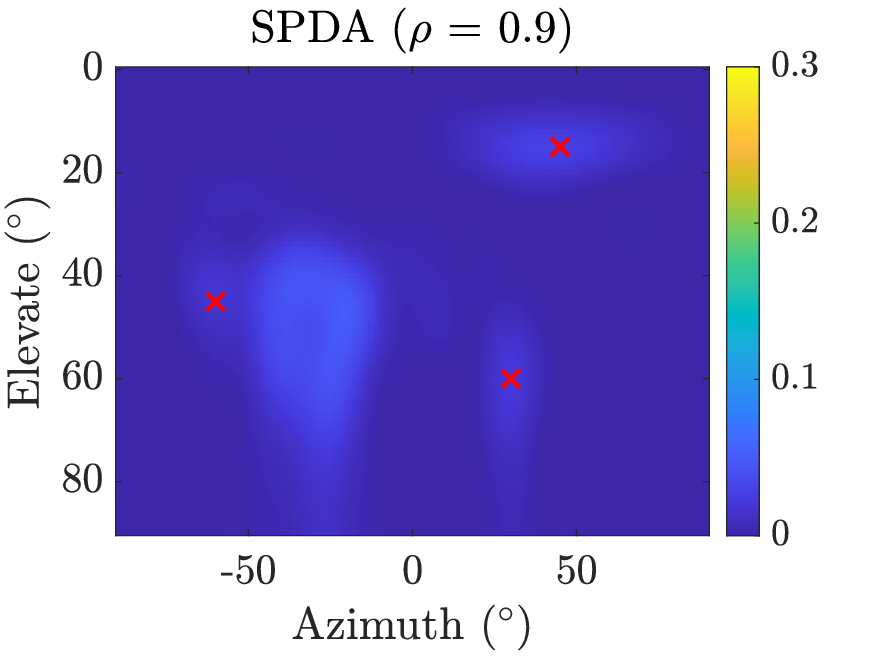}}
        \centering
        \subfigure[]{
		\includegraphics[width=4.2cm]{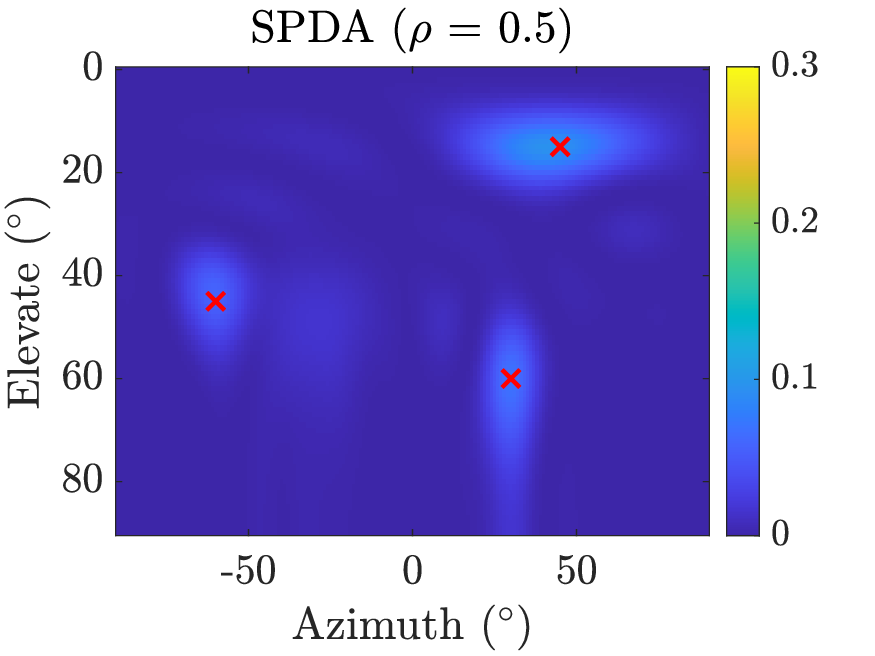}}
        \centering
        \subfigure[]{
		\includegraphics[width=4.2cm]{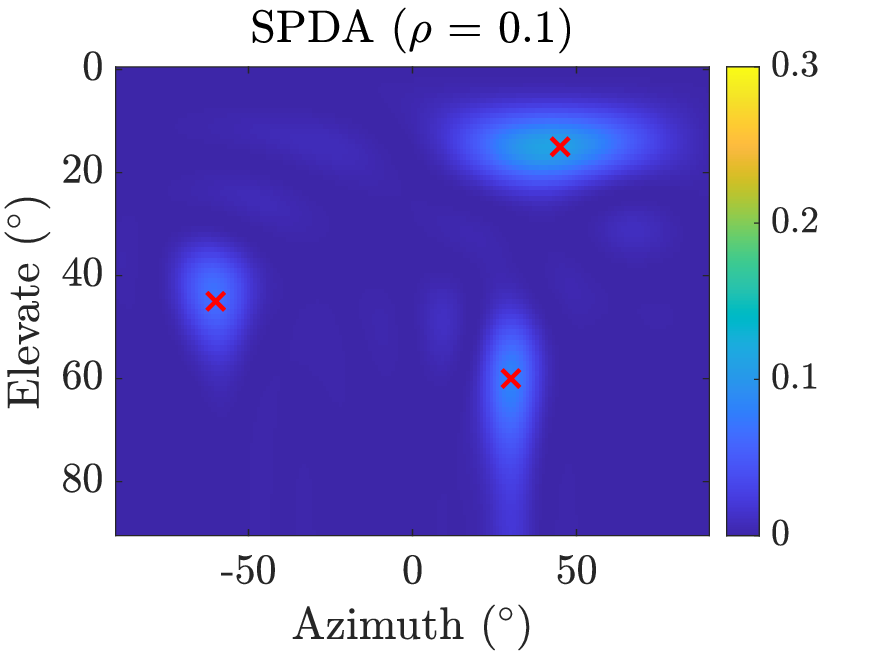}}
        \centering
	\subfigure[]{
		\includegraphics[width=4.3cm]{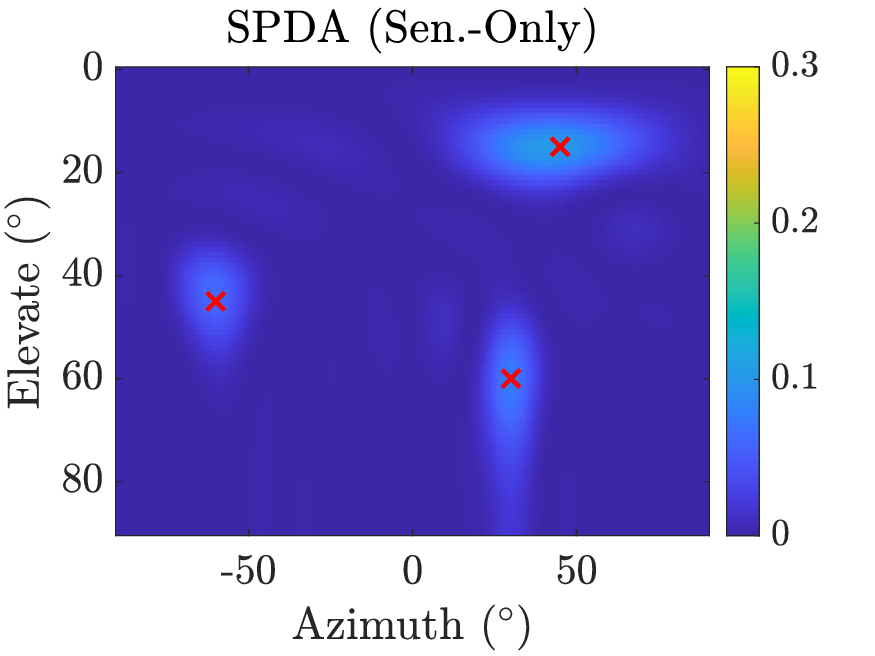}}
	\caption{Beampatterns obtained by different system configurations: (a) CAPA-aided ISAC with $\rho=0.9$. (b) CAPA-aided ISAC with $\rho=0.5$. (c) CAPA-aided ISAC with $\rho=0.1$. (d) CAPA-aided sensing-only system. (e) SPDA-aided ISAC with $\rho=0.9$. (f) SPDA-aided ISAC with $\rho=0.5$. (g) SPDA-aided ISAC with $\rho=0.1$. (h) SPDA-aided sensing-only system.}
	\label{beampattern_fig}		
\end{figure*}
\begin{figure}
	\centering
	\subfigure[]{
		\includegraphics[width=4.2cm]{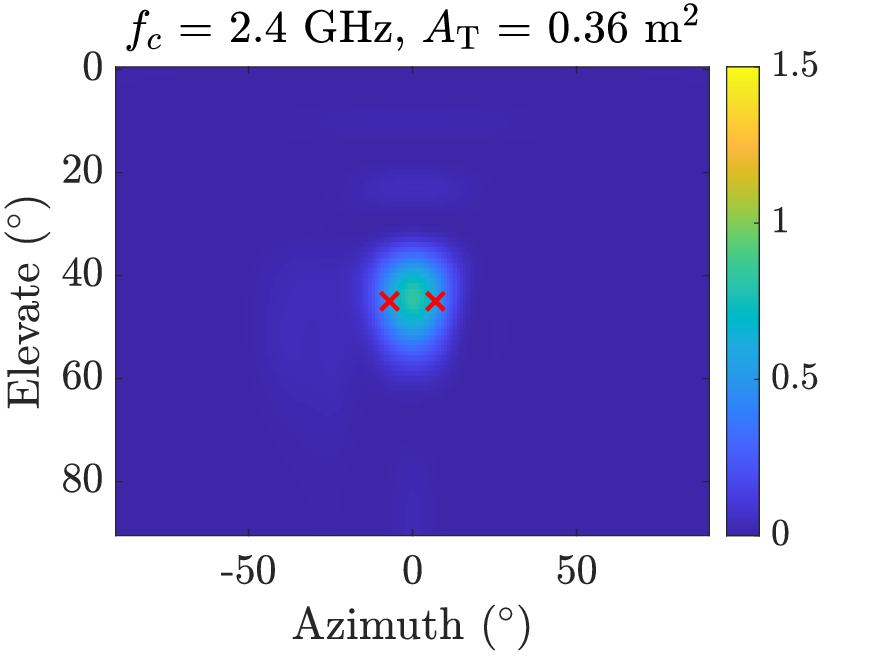}}
	\subfigure[]{
		\includegraphics[width=4.2cm]{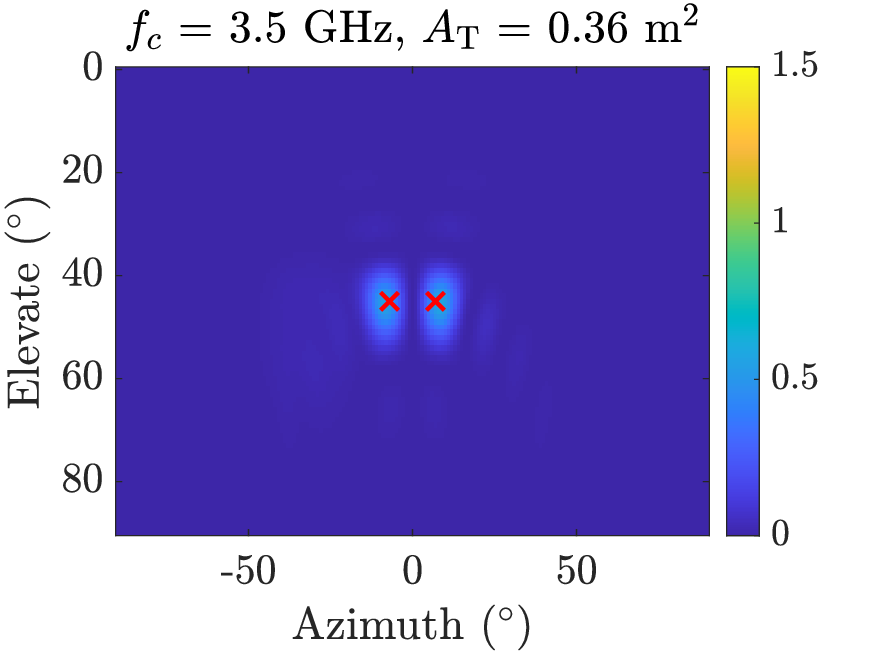}}	
	\centering
	\subfigure[]{
		\includegraphics[width=4.2cm]{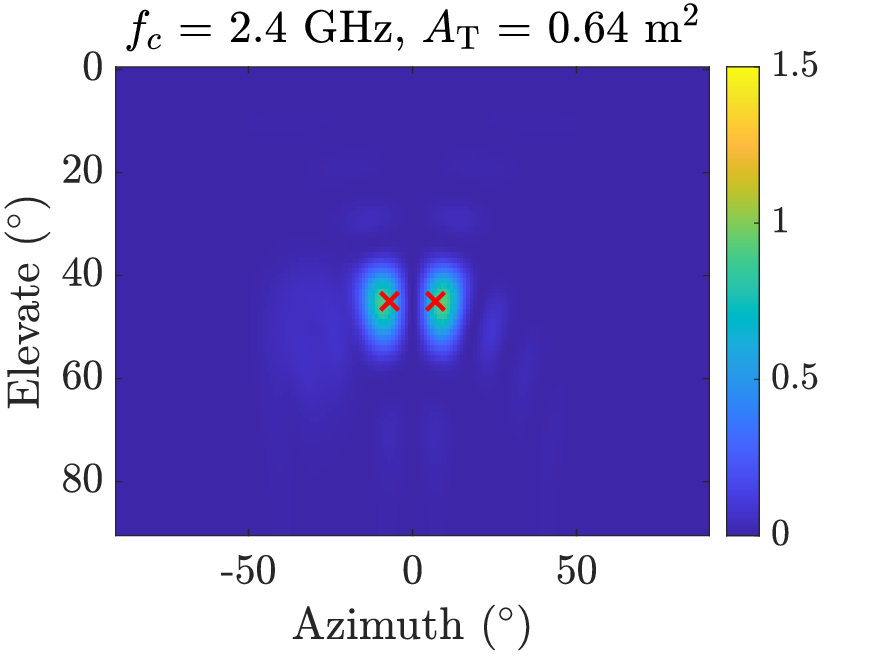}}
	\centering
	\subfigure[]{
		\includegraphics[width=4.2cm]{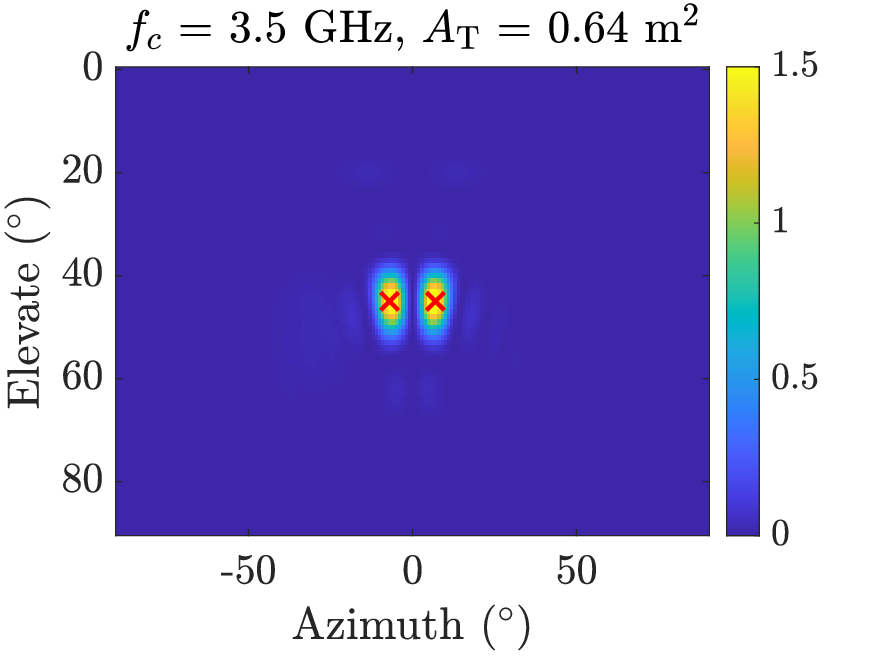}}
	\caption{Beampatterns resolutions under different aperture sizes and carrier frequencies: (a) $f_c=2.4$ GHz and $A_\mathrm{T}=0.36$ m$^2$. (b)  $f_c=3.5$ GHz and $A_\mathrm{T}=0.36$ m$^2$. (c)  $f_c=2.4$ GHz and $A_\mathrm{T}=0.64$ m$^2$. (d)  $f_c=3.5$ GHz and $A_\mathrm{T}=0.64$ m$^2$. }
	\label{resolution_fig}		
\end{figure}
 In this part, the sensing capability of the proposed CAPA-aided ISAC system is examined in terms of the beampattern and the integrated sidelobe-to-mainlobe ratio (ISMR). First, the beampatterns of the CAPA are shown in Fig. \ref{beampattern_fig}. Specifically, Figs. \ref{beampattern_fig}(a)–(c) depict the CAPA beampatterns for $\rho=0.9$, $0.5$, and $0.1$, respectively, while Fig. \ref{beampattern_fig}(d) shows the sensing-only case. It can be observed that for $\rho=0.1$, the beampattern closely resembles that of the sensing-only case, exhibiting three pronounced peaks in the target directions (indicated by red crosses). As $\rho$ increases to $0.5$, the mainlobe gains decrease and the sidelobe levels rise. When $\rho$ further increases, the peaks gradually reduce and the sidelobes eventually become dominant. Figs. \ref{beampattern_fig}(e)–(h) show the corresponding SPDA beampatterns. It can be seen that the beam gains of SPDA are consistently lower than those of CAPA. In particular, the CAPA achieves up to threefold higher beam gains at the target directions, validating its superior beamforming capability compared to the conventional SPDA.

Fig. \ref{resolution_fig} illustrates the beampattern resolutions under different carrier frequencies ($f_c=2.4, 3.5$ GHz) and aperture sizes ($A_\mathrm{T}=0.36, 0.64$ m$^2$). Specifically, two closely spaced targets are considered, located at ($-7^\circ$, $45^\circ$) and ($7^\circ$, $45^\circ$), respectively. It can be observed that when $f_c=2.4$ GHz and $A_\mathrm{T}=0.36$ m$^2$, the two beams merge into a single lobe, indicating limited angular resolution. In contrast, for $f_c=3.5$ GHz and/or $A_\mathrm{T}=0.64$ m$^2$, the two targets can be clearly distinguished with separated mainlobes. Moreover, the beam gain in the case of $f_c=3.5$ GHz and $A_\mathrm{T}=0.64$ m$^2$ is higher than that achieved with either $f_c=3.5$ GHz or $A_\mathrm{T}=0.64$ m$^2$ alone, and significantly better than the low frequency and small aperture case. In addition, the sidelobe level is further reduced under the joint condition of $f_c=3.5$ GHz and $A_\mathrm{T}=0.64$ m$^2$. These results demonstrate that increasing the aperture size and carrier frequency effectively narrows the beamwidth, enhances the mainlobe gain, and suppresses sidelobes. This finding highlights the importance of exploiting higher frequencies and larger apertures to boost sensing capability in future ISAC systems.

 \begin{figure}[!t]
  \centering
    \includegraphics[scale=0.46]{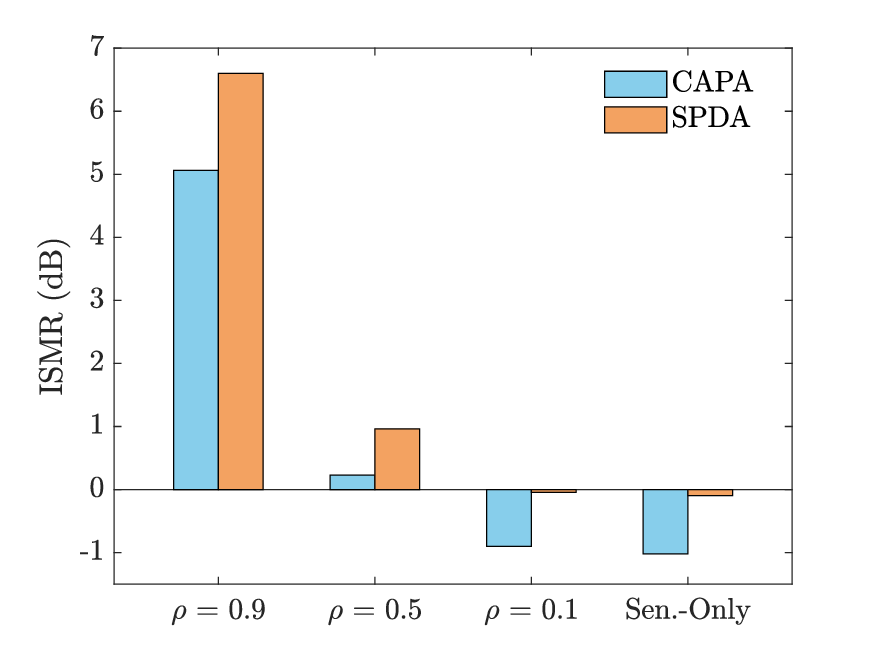}
    \caption{The ISMR (dB) of the CAPA-aided ISAC system under different $\rho$, and comparisons with SPDA.}
    \label{fig:ISMR}
\end{figure}

While the beampattern provides an intuitive visualization of the sensing performance, the ISMR serves as an important quantitative metric that comprehensively reflects the beam gain, sidelobe levels, and energy leakage. In Fig. \ref{fig:ISMR}, the ISMR values of both CAPA and SPDA under different $\rho$ are illustrated. Specifically, the ISMR is defined as the ratio of the integrated power in the sidelobe region to that in the mainlobe region, given by
\begin{align}
\text{ISMR}= \frac{\sum_{{\theta_s,\phi_s}\in \Theta_s } A(j(\mathbf{s}),\theta_s,\phi_s)}{\sum_{{\theta_m,\phi_m}\in \Theta_m } A(j(\mathbf{s}),\theta_m,\phi_m)},
\end{align}
where $A(j(\mathbf{s}),\theta_s,\phi_s)$ is calculated according to (\ref{bp_farfield}), while $\Theta_s$ and $\Theta_m$ denote the sets of sidelobe and mainlobe angles, respectively. Given the target directions of ($45^\circ$, $15^\circ$), ($-60^\circ$, $45^\circ$), and ($30^\circ$, $60^\circ$), the mainlobe region is defined as a square area centered at each target direction with an angular range of $\pm 10^\circ$.
From Fig. \ref{fig:ISMR}, it can be observed that a smaller $\rho$ leads to a lower ISMR, indicating stronger sidelobe suppression. For example, when $\rho=0.1$, the ISMR approaches $-1$ dB, which is close to that of the sensing-only case, whereas for $\rho=0.9$, the ISMR rises to about $5$ dB due to higher sidelobe levels. Moreover, across all cases, CAPA consistently achieves a lower ISMR than SPDA, demonstrating its superior capability in sidelobe suppression. This improvement arises from the additional spatial DoFs provided by the continuous aperture structure of CAPA.

\section{Conclusions} \label{conclusions}
This paper proposed a CAPA-aided multi-user and multi-target ISAC system, in which a joint optimization problem was formulated to minimize communication MUI and sensing beampattern mismatch. To tackle the resultant functional programming problem, a CoV-based algorithm was developed, and the optimal ISAC waveform structure was derived. Numerical results demonstrated that the proposed CAPA design substantially enhances both the sensing beampattern gain and communication BER performance compared to the conventional SPDA, highlighting the effectiveness of CAPA in the application of ISAC.

\begin{appendices}   
\section{Proof of Theorem \ref{theo_optimal_structure}}  \label{app_a}
To prove the theorem \ref{theo_optimal_structure}, CoV can be employed. Assume that  $\overline{j}(\mathbf{s})$ is an optimal solution that minimizes $\mathcal{L} (j(\mathbf{s})),\mu)$, for any $\epsilon \longrightarrow 0$, there is
\begin{align}
    \mathcal{L}(j(\mathbf{s})),\mu)\leq \mathcal{L}(j(\mathbf{s})+\epsilon {\eta}(\mathbf{s}),\mu)\triangleq \mathcal{F}(j(\mathbf{s})),
\end{align}
where $\epsilon {\eta}(\mathbf{s})$ represents a
variation of $j(\mathbf{s})$, and ${\eta}(\mathbf{s})$ is an arbitrary smooth function satisfying ${\eta}(\mathbf{s})=0, \forall \mathbf{s} \in \partial \mathcal{S}$. By treating $\epsilon$ as a real-valued variable, the functional $\mathcal{F}(j(\mathbf{s}))$ can also be regarded as a function of $\epsilon$, defined by $\Phi(\epsilon) \triangleq\mathcal{F}(j(\mathbf{s})) $. In this way, the analysis of the functional $\mathcal{F}(j(\mathbf{s}))$ is reduced to the analysis of the real function $\Phi(\epsilon)$.  Specifically,  the explicit expression form of $\Phi(\epsilon)$ can be given by
\begin{align} \label{Phi_epsilon}
    \Phi(\epsilon) =&   2\rho\mathfrak{R}\left\{\sum_{k=1}^K\int_{\mathcal{S}_\mathrm{T}} \int_{\mathcal{S}_\mathrm{T}} \epsilon {\eta}^*(\mathbf{s}) H^*_k(\mathbf{s})  H_k(\mathbf{s}')  j(\mathbf{s}') d\mathbf{s}' d\mathbf{s}\right\}  \notag\\
    & -  2\rho\mathfrak{R}\left \{\sum_{k=1}^K c_k \int_{\mathcal{S}_\mathrm{T}} \epsilon \eta^*(\mathbf{s}) H_k^*(\mathbf{s})    d\mathbf{s} \right \} \notag\\
    &- 2 (1-\rho)  \mathfrak{R} \left\{\int_{\mathcal{S}_\mathrm{T}} \epsilon  \eta^*(\mathbf{s}) j_{d}(\mathbf{s}) d\mathbf{s}  \right\} \notag \\
    & +\lambda 2 \Re\left\{\int_{\mathcal{S}_\mathrm{T}} \epsilon  \eta^*(\mathbf{s}) j(\mathbf{s})d\mathbf{s} \right\} + \widetilde{\Phi}(\epsilon^2)+C,
\end{align}
where $\widetilde{\Phi}(\epsilon^2)$ collects all the terms w.r.t to $\epsilon^2$ while $C$ represents the sum of terms independent of $\epsilon$. Since $\overline{j}(\mathbf{s})$ is an optimal solution, the minimum of $\mathcal{L}(j(\mathbf{s}),\mu)$ occurs at $\epsilon =0$, which satisfies
\begin{align} \label{CoV_extre_point}
    \left.\frac{d \Phi(\epsilon)}{d \epsilon}\right|_{\epsilon=0}=0.
\end{align}
By substituting (\ref{Phi_epsilon}) to (\ref{CoV_extre_point}), we can construct an equation as follows:
\begin{align} \label{eq_CoV0}
    &\rho\mathfrak{R}\left\{\sum_{k=1}^K\int_{\mathcal{S}_\mathrm{T}} \int_{\mathcal{S}_\mathrm{T}} {\eta}^*(\mathbf{s}) H^*_k(\mathbf{s})  H_k(\mathbf{s}')  j(\mathbf{s}') d\mathbf{s}' d\mathbf{s}\right\}  \notag \\
    &-  \rho\mathfrak{R}\left \{\sum_{k=1}^K c_k \int_{\mathcal{S}_\mathrm{T}}  \eta^*(\mathbf{s}) H_k^*(\mathbf{s})    d\mathbf{s} \right \} \notag\\
    &-  (1-\rho)   \mathfrak{R} \left\{\int_{\mathcal{S}_\mathrm{T}} \eta^*(\mathbf{s}) j_{d}(\mathbf{s}) d\mathbf{s}  \right\} \notag \\
    &+\mu  \Re\left\{\int_{\mathcal{S}_\mathrm{T}}   \eta^*(\mathbf{s}) j(\mathbf{s})d\mathbf{s} \right\} = 0,
\end{align}
which implies the optimal $j(\mathbf{s})$ should comply with (\ref{eq_CoV0}).
For better inspection, it can be further simplified as
\begin{align} \label{eq_CoV1}
    & \mathfrak{R}\left\{  \int_{\mathcal{S}_\mathrm{T}} {\eta}^*(\mathbf{s}) v(\mathbf{s})  d\mathbf{s}\right\}   = 0,
\end{align}
where we denote
\begin{align}
    v(\mathbf{s}) &= \rho \sum_{k=1}^K H^*_k(\mathbf{s})\int_{\mathcal{S}_\mathrm{T}}   H_k(\mathbf{s}')  j(\mathbf{s}') d\mathbf{s}' \notag \\
    &-  \rho \sum_{k=1}^K c_k H^*_k(\mathbf{s}) -  (1-\rho) j_{d}(\mathbf{s}) + \mu  j(\mathbf{s}).
\end{align}
Considering that (\ref{eq_CoV1}) must be satisfied for any arbitrary
function ${\eta}(\mathbf{s})$, $v(\mathbf{s})=0$ must hold according to the \textbf{Lemma \ref{lem_cov}}, which yields
\begin{align}
     \mu j(\mathbf{s}) &= -\rho \sum_{k=1}^K H^*_k(\mathbf{s})\int_{\mathcal{S}_\mathrm{T}}   H_k(\mathbf{s}')  j(\mathbf{s}') d\mathbf{s}'  \notag \\
    & +  \rho \sum_{k=1}^K c_k H^*_k(\mathbf{s})  + (1-\rho) j_{d}(\mathbf{s}).
\end{align}
 This completes the proof.

\section{Proof of Theorem \ref{theo_lambda_eq}}  \label{app_b}
While $j(\mathbf{s})$ with the structure of (\ref{optimal_structure2}) can achieve minimal objective function, the optimal $\mu^\star$ should also guarantee the power constraint (\ref{ISAC_formulation}{b}), i.e.,
\begin{align} \label{power_constraint}
    \int_{\mathcal{S}_\mathrm{T}} j(\mathbf{s}) j^*(\mathbf{s})d\mathbf{s} & = P_t.
\end{align}
For the convenience of the subsequent derivations, we multiply both sides of (\ref{power_constraint}) by $\mu$:
\begin{align}
    \mu^2 \int_{\mathcal{S}_\mathrm{T}} j(\mathbf{s}) j^*(\mathbf{s})d\mathbf{s} & = \mu^2 P_t.
\end{align}

\begin{figure*}[!t] 
\begin{align} \label{lambda_j_square}
    &|\mu   j(\mathbf{s}) |^2=\mu^2   j(\mathbf{s}) j^*(\mathbf{s})  \notag \\
     =& \rho^2   \sum_{k=1}^K \sum_{k'=1}^K z_{k'}^* H^*_k(\mathbf{s}) H_{k'}(\mathbf{s})  z_k +(1-\rho)^2 |j_{d}(\mathbf{s})|^2 + \rho^2 \sum_{k=1}^K \sum_{k'=1}^K    c_k H^*_k(\mathbf{s}) H_{k'}(\mathbf{s}) c_{k'}^*  \notag  \\
    - & 2 \rho^2 \Re\left\{\sum_{k=1}^K \sum_{k'=1}^K z_k H^*_k(\mathbf{s})   H_{k'}(\mathbf{s}) c_{k'}^*\right\} -2 \rho (1-\rho) \Re\left\{\sum_{k=1}^K z_k H^*_k(\mathbf{s})  j_{d}^*(\mathbf{s}) \right\} + 2 \rho (1-\rho) \Re\left\{\sum_{k=1}^K c_k H^*_k(\mathbf{s}) j_{d}^*(\mathbf{s}) \right\}.
\end{align} 
\hrulefill
\begin{align} \label{lambda_j_square_integral}
    \mu^2   \int_{\mathcal{S}_\mathrm{T}} j(\mathbf{s}) j^*(\mathbf{s})  d\mathbf{s}
    &= \rho^2   \sum_{k=1}^K \sum_{k'=1}^K z_{k'}^* z_k \int_{\mathcal{S}_\mathrm{T}} H^*_k(\mathbf{s}) H_{k'}(\mathbf{s})   d\mathbf{s}  + \rho^2 \sum_{k=1}^K \sum_{k'=1}^K    c_k c_{k'}^* \int_{\mathcal{S}_\mathrm{T}} H^*_k(\mathbf{s})  H_{k'}(\mathbf{s})d\mathbf{s} \notag \\
    & +(1-\rho)^2 \int_{\mathcal{S}_\mathrm{T}} j_{d} (\mathbf{s}) j_{d}^*(\mathbf{s}) d\mathbf{s}  -2 \rho^2 \Re\left\{\sum_{k=1}^K \sum_{k'=1}^K z_k c_{k'}^* \int_{\mathcal{S}_\mathrm{T}}  H^*_k(\mathbf{s})  H_{k'}(\mathbf{s}) d\mathbf{s}\right\}  \notag \\
    &-2 \rho (1-\rho) \Re\left\{\sum_{k=1}^K z_k^* \int_{\mathcal{S}_\mathrm{T}}  H_k(\mathbf{s})  j_{d}(\mathbf{s}) d\mathbf{s} \right\}  + 2 \rho (1-\rho) \Re\left\{\sum_{k=1}^K c^*_k \int_{\mathcal{S}_\mathrm{T}} H_k(\mathbf{s}) j_{d}(\mathbf{s}) d\mathbf{s}  \right\} . 
\end{align}
\hrulefill
\end{figure*}
To explicitly express the term on the left-hand, the optimal structure in (\ref{optimal_structure2}) is substituted to guarantee the optimality. Specifically, the absolute square of $\mu j(\mathbf{s})$ can be expressed as (\ref{lambda_j_square}) on the top of this page. Then, integrate it over $d\mathbf{s}$ and yields (\ref{lambda_j_square_integral}). It can be observed that the integrals $\int_{\mathcal{S}_\mathrm{T}} H^*_k(\mathbf{s}) H_{k'}(\mathbf{s})   d\mathbf{s}$ and $\int_{\mathcal{S}_\mathrm{T}} H_k(\mathbf{s}) j_{d}(\mathbf{s}) d\mathbf{s}$ correspond to $q_{k',k}$ and $u_{k}$, respectively. Therefore, it can be rewritten as
\begin{align}
    & \rho^2   \sum_{k=1}^K \sum_{k'=1}^K z_{k'}^* z_k  q_{k',k}+ \rho^2 \sum_{k=1}^K \sum_{k'=1}^K    c_k c_{k'}^* q_{k',k} \notag \\
    &+ (1-\rho)^2 P_t -2 \rho^2 \Re\left\{\sum_{k=1}^K \sum_{k'=1}^K z_k c_{k'}^* q_{k',k}\right\} \notag \\
    & \!-2 \rho (1\!-\!\rho) \Re\left\{\sum_{k=1}^K z_k^* u_k \right\} \!+\! 2 \rho (1\!-\!\rho) \Re\left\{\sum_{k=1}^K c^*_k u_k \right\},
\end{align}
which can be further simplified in a matrix form, i.e.,
\begin{align}
&\mu ^2   \int_{\mathcal{S}_\mathrm{T}} j(\mathbf{s}) j^*(\mathbf{s})  d\mathbf{s}\notag\\
    =& \rho^2  \mathbf{z}^\mathrm{H} \mathbf{Q} \mathbf{z} + \rho^2 \mathbf{c}^\mathrm{H} \mathbf{Q} \mathbf{c}  + (1-\rho)^2 P_t \notag \\
  &\!\!\!\!\!- 2 \rho^2 \Re\left\{ \mathbf{c}^\mathrm{H} \mathbf{Q} \mathbf{z} \right\}  \!-\!2 \rho (1\!-\!\rho) \Re\left\{ \mathbf{z}^\mathrm{H} \mathbf{u}   \right\} \! + \! 2 \rho (1 \! -\!\rho) \Re\left\{\mathbf{c}^\mathrm{H} \mathbf{u}  \right\} \notag\\
  = &  \rho^2  \mathbf{z}^\mathrm{H} \mathbf{Q} \mathbf{z}  \!-\! 2 \rho^2 \Re\left\{ \mathbf{c}^\mathrm{H} \mathbf{Q} \mathbf{z} \right\}  -2 \rho (1-\rho) \Re\left\{ \mathbf{z}^\mathrm{H} \mathbf{u}   \right\} + \widetilde{c}.
\end{align}
Here, $\widetilde{c}= \rho^2 \mathbf{c}^\mathrm{H} \mathbf{Q} \mathbf{c} \!+\! (1\!-\!\rho)^2 P_t \!+\! 2 \rho (1\!-\!\rho) \Re\left\{\mathbf{c}^\mathrm{H} \mathbf{u}  \right\}$ collects
all the terms that are unrelated to $\mathbf{z}$. Meanwhile, $\mathbf{z}$ is a function of $\mu$, as expressed in (\ref{z_expression_mat}). Consequently, the optimal $\mu$ that satisfies the power constraint and the optimality of the objective function can be obtained by solving the following equation:
    \begin{align} 
    &\rho^2   \mathbf{z}^\mathrm{H} (\mu) \mathbf{Q} \mathbf{z}(\mu)  \! -\!\!2 \rho^2 \Re\left\{ \mathbf{c}^\mathrm{H} \mathbf{Q} \mathbf{z} (\mu) \right\}  \notag \\
    &\!-\!\!2 \rho (1\!-\!\rho) \Re\left\{ \mathbf{z}(\mu)^\mathrm{H} \mathbf{u}   \right\} \!+\! \widetilde{c} \!= \! \mu^2 P_t.
\end{align}
The proof ends.
\end{appendices}

\bibliographystyle{IEEEtran}
\bibliography{references2}{}

\end{document}